\documentclass[fleqn,10pt,table]{wlscirep}
\usepackage[T1]{fontenc}

\usepackage{amsmath,amsbsy,epsfig,epstopdf,amssymb}
\usepackage{dsfont}
\usepackage{xcolor}
\usepackage[normalem]{ulem}
\usepackage{makecell}
\usepackage{multirow}

\usepackage{xcolor, graphicx}
\usepackage{multirow}
\usepackage{mathtools}
\usepackage{ulem}
\usepackage[autostyle]{csquotes}
\usepackage{comment}
\usepackage{algorithm} 
\usepackage{algorithmic} 
\usepackage{pgfplots}
\usepackage{graphicx}
\usepackage{subcaption}
\usepackage{mwe}
\usepackage{adjustbox}
\usepackage{xcolor}
\usepackage{collcell}
\usepackage{pgfplots}
\pgfplotsset{compat=1.17} 
\usepackage{pgf-pie}
\usepackage{etoolbox}

\definecolor{White}{HTML}{FFFFFF}
\definecolor{LightGray}{HTML}{dee2e6}

\usepackage{caption}
\usepackage{subcaption}

\graphicspath{{./img/}}

\title{Experimental implementation of a neural network optical channel equalizer in restricted hardware using pruning and quantization}

\author[1, *]{Diego Arg\"uello Ron}
\author[1,2]{Pedro J. Freire}
\author[1]{Jaroslaw E. Prilepsky}
\author[1]{Morteza Kamalian-Kopae}
\author[2]{Antonio Napoli}
\author[1, \dag]{Sergei K. Turitsyn}

\affil[1]{Aston Institute of Photonic Technologies, Aston University, B4 7ET, Birmingham, UK}  
\affil[2]{Infinera, Sankt-Martin-Str. 76, 81541, Munich, Germany} 

\affil[*]{d.arguelloron@aston.ac.uk, ${}^\dag$s.k.turitsyn@aston.ac.uk}

\begin{abstract}
The deployment of artificial neural networks-based optical channel equalizers on edge-computing devices is critically important for the next generation of optical communication systems. However, this is still a highly challenging problem, mainly due to the computational complexity of the artificial neural networks (NNs) required for the efficient equalization of nonlinear optical channels with large dispersion-induced memory. To implement the NN-based optical channel equalizer in hardware, a substantial complexity reduction is needed, while we have to keep an acceptable performance level of the simplified NN model. In this work, we address the complexity reduction problem by applying pruning and quantization techniques to an NN-based optical channel equalizer. We use an exemplary NN architecture, the multi-layer perceptron (MLP), to mitigate the impairments for 30~GBd 1000~km transmission over a standard single-mode fiber, and demonstrate that it is feasible to reduce the equalizer’s memory by up to 87.12\%, and its complexity by up to 78.34\%, without noticeable performance degradation. In addition to this, we accurately define the computational complexity of a compressed NN-based equalizer in the digital signal processing (DSP) sense. Further, we examine the impact of using different CPU and GPU settings on the power consumption and latency for the compressed equalizer. We also verify the developed technique experimentally, by
implementing the reduced NN equalizer on two standard edge-computing hardware units: Raspberry Pi 4 and Nvidia Jetson Nano, which are used to process the data generated via simulating the signal's propagation down the optical-fiber system.

\end{abstract}

\begin{document}

\flushbottom
\maketitle
%
%
\thispagestyle{empty}

\section*{Introduction}
Optical communications form the backbone of the global digital infrastructure. Nowadays, optical networks are the main providers of global data traffic, not only interconnecting billions of people, but also supporting the life-cycle of a huge number of different autonomous devices, machines, and control systems. One of the major factors limiting the throughput of contemporary fiber-optic communication systems is the nonlinearity-induced transmission impairments \cite{Winzer:18,Cartledge:17}, emerging from both the fiber media's nonlinear response and the system's components. The existing and potential solutions to this problem include, e.g., the mid-span optical phase conjugation, digital back-propagation (DBP), and inverse Volterra series transfer function, to mention just a few noticeable methods\cite{Rafique,Cartledge:17,Dar}.
But, it should be stressed that in the telecommunication industry, the competition between possible solutions occurs not only in terms of performance but also in terms of hardware deployment options, operational costs, and power consumption. 

During the last years, the approaches based on machine learning techniques and, in particular, those utilizing NNs, have become an increasingly popular topic of research, as the NNs can efficiently unroll both fiber and component-induced impairments\cite{Zibar02,nevin2021machine,Jar, hager2018nonlinear,zhang2019field,freire2021performance,del2020,deligiannidis2021performance,PedroOFC,sidelnikov2018,sidelnikov2019methods}. One of the straightforward ways of using a NN for signal's corruption compensation in optical transmission systems is to plug it into the system as a post-equalizer \cite{Jar,freire2021performance,sidelnikov2018}, a special signal processing device at the receiver side, aimed at counteracting the detrimental effects emerging during the data transmission \cite{barry2012digital}. Numerous preceding studies have demonstrated the potential of this type of solution \cite{Jar,hager2018nonlinear}. A number of NN architectures have already been analyzed in different types of optical systems (submarine, long-haul, metro, and access). These architectures include the feed-forward NN designs such as the MLP \cite{Jar,sidelnikov2018,freire2021performance,sidelnikov2019methods}, considered in the current study, or more sophisticated recurrent-type NN structures \cite{freire2021performance,del2020,deligiannidis2021performance, ming2021ultralow}. However, the practical deployment of real-time NN-based channel equalizers implies that their computational complexity is, at least, comparable, or, desirably lower than that of existing conventional digital signal processing (DSP) solutions \cite{Kaneda:20}, and remains a matter of debate. This is a relevant aspect because the good performance achieved by the NNs is typically linked to the use of a large number of parameters and floating-point operations \cite{freire2021performance}. The high computational complexity leads, in turn, to high memory and computing power requirements, increasing the energy and resource consumption \cite{blalock2020state,han2016deep}. Thus, the use of NN-based methods, while being, undoubtedly, promising and attractive, faces a major challenge in optical channel equalization, where the computational complexity emerges as an important limiting real-time deployment factor \cite{han2016deep, srinivas2017training, freire2021performance,deligiannidis2021performance}. We notice here that it is, of course, well known that some NN architectures can be simplified without significantly affecting their performance, thanks, e.g., to strategies such as pruning and quantization \cite{blalock2020state, han2016deep, hawks2021ps,sze2017efficient, liang2021pruning,fujisawa2021}. However, their application in the experimental environment of the resource-restricted hardware has not been fully studied \textit{in the context of coherent optical channel equalization} yet. It is also necessary to understand and further analyze the trade-off between the complexity reduction and the degradation of system performance, as well as the complexity reduction impact on the end device's energy consumption.

In this paper, we apply the pruning and quantization techniques to reduce the hardware requirements of an NN-based coherent optical channel equalizer, while keeping its performance at a high level. We also emphasize the importance of an accurate evaluation of the equalizer's computational complexity in the DSP sense. Apart from the complexity and inference time study, an additional novelty and advance of our work lie in the energy consumption analysis and the study of the impact that the characteristics of both the hardware and the model, have on these metrics.

\section*{Results}
We develop and experimentally evaluate the performance of a low-complexity NN-based equalizer that can be deployed on resource-constrained hardware and, at the same time, can successfully mitigate nonlinear transmission impairments in a simulated optical communication system. This is achieved by applying the pruning and quantization techniques to the NN \cite{sze2017efficient}, and by studying the optimal trade-off between the complexity of the NN solution and its performance. The obtained results can be split into three main categories.  

First, we quantify how complexity reduction techniques affect the performance of the NN model and establish a compression limit for optimal performance versus complexity trade-off. Second, we analyze the computational complexity of the pruned and quantized NN-based equalizer in terms of DSP. Finally, we experimentally evaluate the impact that the characteristics of the hardware and the NN model have on the signal processing time and energy consumption by deploying the latter on both a Raspberry Pi 4 and an Nvidia Jetson Nano.

Now we briefly review the previous results in the field of compression techniques applied to NN-based equalizers in optical links, to underline the novelty of our current approach. The use of these techniques to reduce the NNs complexity in optical systems is, clearly, not a new concept\cite{fujisawa2021}. However, the compression methods have recently gained a new wave of attention due to the question of how realistic the hardware implementation of NN-based equalizers in optical transmission systems is. In a direct detection transmission system, a parallel-pruned NN equalizer for a 100-Gbps PAM-4 links were tested experimentally using the enhanced version of the one-shot pruning method\cite{li2021high}, which decreased by 50\% the resource consumption without significant performance degradation. When considering coherent optical transmission, the complexity of the so-called learned DBP nonlinearity mitigation method was reduced by pruning the coefficients in the finite impulse response filters\cite{oliari2020revisiting} (see more technical explanations in the Methods section below).
In that case, using a cascade of three filters, a sparsity level of around 92\% can be achieved with a negligible impact on the overall performance. Recently, some advanced techniques for avoiding multiplications in such equalizers using additive powers-of-two quantization were tested \cite{Koike2021}. In the latter work, the 99\% percent of weights can be removed using advanced pruning techniques, and instead of multiplications, just bit-shift operations are required.
However, none of those works deal with the experimental demonstration of hardware implementation, and our study addresses exactly the latter problem.

So, unlike previous works, in the current study, we implement the compressed NN-based equalizer for the coherent optical channel in two different hardware platforms: a Raspberry Pi 4 and an Nvidia Jetson Nano. We also evaluate the impact of the compression techniques on the system's latency for each hardware type and study the performance-complexity trade-off. Finally, we carry out an analysis of energy consumption and of the impact that the characteristics of the hardware and the NN model have on it.

\subsection*{Optical Communication System and Equalizer Design}\label{Sec:EqDesign}

\begin{figure}[!ht]
    \centering
    \includegraphics[width=17.5cm, height = 3.5cm]{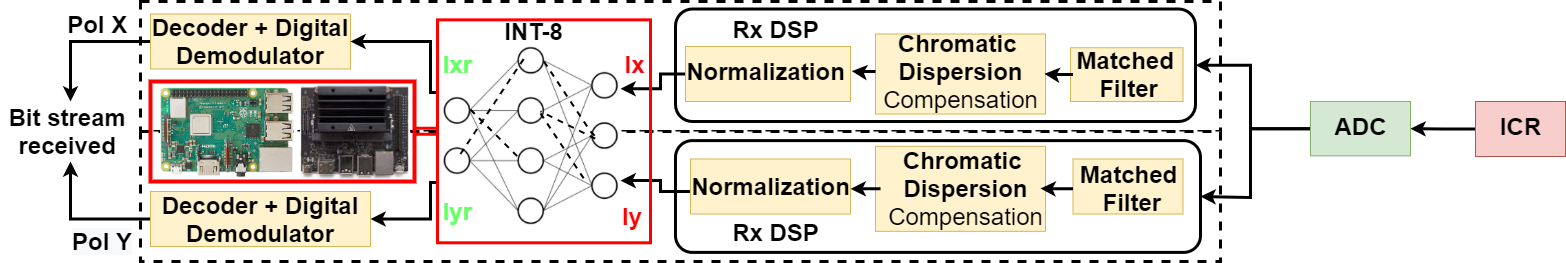}
    \caption{Structure of a communication channel that is equalized using a pruned and quantized neural network deployed on resource-restricted hardware (e.g. a Raspberry Pi 4 or a Nvidia Jetson Nano). }
    \label{fig:figure1}
\end{figure}

\begin{figure*}[t!] 
\centering
\begin{tikzpicture}[scale=0.86]
\begin{axis} [
xlabel={Launch power [dBm]},
ylabel={Q-Factor [dB]},
grid=both,
xmin=-3, xmax=2,
xtick={-3, ..., 5},
ymax =9,
legend style={legend pos=south west, legend cell align=left,fill=white, fill opacity=0.6, draw opacity=1,text opacity=1},
grid style={dashed}]
]

\addplot[color=blue, mark=triangle, dashed, very thick] coordinates {
(-3,7.53)(-2,8.04)(-1,8.27)(0,8.19)(1,7.85)(2,7.04)
};
\addlegendentry{\footnotesize{MLP (Test)}};

\addplot[color=green, very thick] coordinates {
(-3,7.46)(-2,7.74)(-1,7.77)(0,7.40)(1,6.60)(2,5.59)

};
\addlegendentry{\footnotesize{Regular DSP}};

\end{axis}
\end{tikzpicture}
\caption{Performance comparison for the NN-based equalizer with respect to the Regular DSP.}
\label{fig:OptimalPower}
\end{figure*}
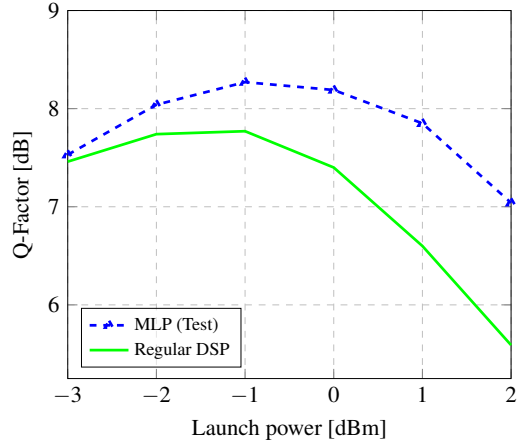

To address the use of an MLP as an NN-based equalizer, an accurate measurement system for both the inference time and the power consumption, on both Raspberry Pi and Nvidia Jetson Nano, has been designed, so that the effects that pruning and quantization have on these metrics, can be characterized (see the Methods section below for a detailed explanation). In Refs.~\cite{sidelnikov2018,freire2021performance}, the non-compressed MLP post-equalizer was considered, and it was shown that it can successfully compensate for the nonlinearity-induced impairments in a coherent optical communication system. We analyze the equalizer's performance in terms of the standard achieved Q-factor, using the simulated data for a 0.1 root-raised cosine (RRC) dual-polarization signal, with 30 GBd, and 64-QAM modulation, for the transmission over the 20$\times$50 km links of standard single-mode fiber (SSMF). We used the same simulator as described in Refs.~\cite{freire2021performance,freire2021transfer}, to generate our training and testing datasets, and the same procedure to training the NN-based equalizer (see Numerical Setup and Neural Network Model subsection in Methods for more details).  In our configuration, the NN is placed at the receiver (Rx) side after the Integrated Coherent Receiver (ICR), Analog-Digital Converter (ADC), and DSP block. This last block consists of a matched filter and a linear equalizer. Concerning the matched filter, it is just the same  RRC filter used in the transmitter. Moreover, the linear equalizer is composed of a full electronic chromatic dispersion compensation (CDC) stage and a normalization step, see Fig.~\ref{fig:figure1}. The CDC uses a frequency-domain equalizer and downsampling to the symbol rate, followed by a phase/amplitude normalizer to the transmitted ones. This normalization process  can be viewed as its normalization by a constant $K_\text{DSP}$ learned using the followed equation:

\begin{equation}
    \label{eqn:norming}
    \mathcal{K}_\text{DSP} = \min_\mathcal{K}\left\|\mathcal{K}\cdot x_{h\!/\!v}(z,t) - x_{h\!/\!v}(0,t)\right\|,
\end{equation}

where constants $K, \, K_\text{DSP} \in \mathbb{C}$, $x_{h\!/\!v}$ is the signal in either $h$ or $v$ polarisation. No other distortions -- related to the components within the transceiver -- were considered.
\\
 For this system, the best optimal power occurred at -1 dBm with the Q-factor being close to 7.8, as it can be appreciated in Fig. \ref{fig:OptimalPower}. We then wanted to investigate the 3 next powers (e.g. 0 dBm, 1 dBm, and 2 dBm) going towards the higher nonlinear regime, where the task of the NN would be more complicated.

The hyperparameters that define the structure of the NN are obtained using a Bayesian optimizer (BO) \cite{freire2021performance,pelikan1999boa}, where the optimization is carried out with regards to the signal's restoration quality performance (see ``Numerical Setup and Neural Network Model'' subsection in Methods). The resulting optimized MLP has three hidden layers (we did not optimize the number of layers, but only the number of neurons and the activation functions type), with 500, 10, and 500 neurons, respectively. (These numbers were set as the minimal and maximal wights number limits, within which the Bayesian was searching the optimal configuration). The activation function ``$\tanh$'' was chosen by the optimizer and no bias is employed. The NN takes the downsampled signal (1 sample per symbol) and inputs into the equalizer $N = 10$ neighbors symbols (number of taps) to recover the central one. This memory size was defined by the BO procedure. The NN was subjected to pruning and quantization after it had been trained and tested. We analyzed the performance of different NN models depending on their sparsity level; the latter ranged from $20\%$ to $90\%$, with a $10\%$ increment. The weights and activations are quantized, converting their data type from 32-bit single-precision floating-point (FP32) to 8-bit integer (INT8). The quantization was carried out to enable a real-time use of the model as well as its deployment on resource-constrained hardware. The final system is depicted in Fig.~\ref{fig:figure1}. The inference process (the signal equalization) was, first, carried out using a MSI GP76 Leopard personal computer, equipped with  Intel\textsuperscript{\tiny\textregistered} Core\textsuperscript{TM} i9-10870H processor, 32GB of RAM and GPU Nvidia RTX2070. The results obtained on this computer were used as a benchmark and compared to those obtained on two small single-board computers: Raspberry Pi 4 and Nvidia Jetson Nano.

Finally, the NNs were developed using TensorFlow. The pruning and quantization techniques were implemented using the TensorFlow Model Optimization Toolkit -- Pruning API and TensorFlow Lite \cite{tensorflow2015-whitepaper}. 

\subsection*{Compressing Process for Neural Network Equalizers}

When designing a NN for a particular purpose, the traditional approach consists in using dense and over-parametrized models, insofar as it often can provide a good model's performance and learning capabilities \cite{hoefler2021sparsity,allen2019convergence}. This is due to the over parametrization's smoothing effect on the loss function, which benefits the convergence of the gradient descent techniques used to optimize the model \cite{hoefler2021sparsity}. However, some precautions must be taken while training an over-parametrized model, because such models often tend to overfit, and their generalization capability can be degraded \cite{hoefler2021sparsity, neill2020overview}.

The good performance achieved due to over-parameterization comes at the cost of larger computational and memory resources. This also results in a longer inference time (latency growth) and higher energy consumption. Note that these costs are the consequence of parameter redundancy and a large number of floating-point operations \cite{han2016deep,sze2017efficient}. Therefore, the capabilities of high-complexity NN-based equalizers do not translate yet into end-user applications on resource-constrained hardware. Thus, reducing the gap between the algorithmic solutions and the experimental real-world implementations is an increasingly active topic of research.
During the past several years, noticeable efforts have been invested in developing techniques that can help to simplify the NNs without significantly decreasing their performance. These techniques are grouped under the term "NNs compression methods", and the most common approaches there, are: down-sizing the models, factorizing the operators, quantization, parameter sharing or pruning \cite{han2016deep, liang2021pruning, sze2017efficient}. When these techniques are applied, the final model typically becomes much less complex, and, therefore, its latency, or the time it takes to make a prediction, decreases, which also results in a lower energy consumption \cite{han2016deep}. In this work, we focus on both pruning and quantization for compressing our NN equalizer and quantify a trade-off between complexity reduction and system performance, see the Methods section for a detailed description of both approaches.

 \subsection*{Performance vs. Compression Trade-off}
\begin{figure}[ht]
\centering
\begin{subfigure}{.5\textwidth}
  \centering
  \begin{tikzpicture}[scale=0.90]
      \begin{axis} [
        xlabel={Sparsity [\%]},
        ylabel={Q-factor [dB]},
        grid=both,   
        xmin=20, xmax=90,
        xtick=data,
        legend style={nodes={scale=0.75, transform shape},legend pos=south west, legend cell align=left,fill=white, fill opacity=0.6, draw opacity=1,text opacity=1},
        grid style={dashed}]
        ]

	\addlegendentry{\footnotesize{NN in RB/Nano/PC 0 dBm}};
    \addplot[color=blue, mark = star, dotted, very thick]         coordinates {
         (20,7.991247436299087)(30, 8.046674328334415)(40,8.098567719841627)(50,8.122927886202735)(60,8.079996047540652)(70,8.002597152571857)(80,7.63715452302044)(90,6.914238885083767)};
		 
		 \addlegendentry{\footnotesize{NN in RB/Nano/PC 1 dBm}};
    \addplot[color=red, mark = star, dotted, very thick]         coordinates {
         (20,7.65397292125604)(30,7.590060606019913)(40,7.692400202429108)(50,7.60726963006983)(60,7.572203816701735)(70,7.493913291492271)(80,7.118207793490187)(90,6.409370165103789)};

		 \addlegendentry{\footnotesize{NN in RB/Nano/PC 2 dBm}};
    \addplot[color=green, mark = star, dotted, very thick]         coordinates {
         (20,6.853854183567501)(30,6.904350919235306)(40,6.903190514435536)(50,6.84535874944205)(60,6.70768282338051)(70,6.605519694011882)(80,6.15713127889077)(90,5.242227295126436)};

\addlegendentry{\footnotesize{LE 0 dBm}};
\addplot [domain = 20:90,
        dotted,
        blue,
        very thick
        ]{7.357567016749218};		
\addlegendentry{\footnotesize{LE 1 dBm}};
\addplot [domain = 20:90,
        dotted,
        red,
        very thick
        ]{6.594191476776125};		
\addlegendentry{\footnotesize{LE 2 dBm}};
\addplot [domain = 20:90,
        dotted,
        green,
        very thick
        ]{5.5706119177563774};	
\addplot [domain = 20:90,
        thick,
        blue,
        very thick
        ]{8.198940145352298};		
\addplot [domain = 20:90,
        thick,
        red,
        very thick
        ]{7.850830980822475};		
\addplot [domain = 20:90,
        thick,
        green,
        very thick
        ]{7.0391262602473645};	

    \end{axis}
    \end{tikzpicture}
  \caption{}
  \label{fig:figure2} 
\end{subfigure}%
\begin{subfigure}{.5\textwidth}
  \centering
  \begin{tikzpicture}[scale=0.9]
      \begin{axis} [
        xlabel={Sparsity [\%]},
        ylabel={Q-factor [dB]},
        grid=both,   
        xmin=20, xmax=90,
        xtick=data,
        legend style={legend pos=south west, legend cell align=left,fill=white, fill opacity=0.6, draw opacity=1,text opacity=1},
        grid style={dashed}]
        ]
            \addlegendentry{\footnotesize{Only Pruned}};
    \addplot[color=blue, mark = star, dashed, very thick]         coordinates {
         (20,7.806504424800339)(30,7.832261050370741)(40,7.8510673427197)(50,7.827591602594758)(60,7.761492104756682)(70,7.6134552478705055)(80,7.236385543381384)(90,6.571817486666757)};
		 
		 \addlegendentry{\footnotesize{Pruned and Quantized}};
    \addplot[color=red, mark = star, dashed, very thick]         coordinates {
         (20,7.679499181777112)(30,7.62194307466425)(40,7.661380640402044)(50,7.656297774692781)(60,7.602541496511819)(70,7.496044296505254)(80,7.104634427676681)(90,6.401361915047229)};

\addlegendentry{\footnotesize{Original}};
\addplot [domain = 20:90,
        thick,
        blue,
        very thick
        ]{7.792969063163775};
\addlegendentry{\footnotesize{Only Quantized}};
\addplot [domain = 20:90,
        thick,
        red,
        very thick
        ]{7.52909238209432};
    \end{axis}
    \end{tikzpicture}
  \caption{ }
  \label{fig:Qfact_sparsity_vs_quantization} 
\end{subfigure}
\caption{a) Q-factor achieved for pruned and quantized models versus the level of sparsity for datasets corresponding to three launch powers: 0 dBm, 1 dBm, and 2 dBm; The solid lines correspond to the Q-factor achieved by the original model. The dashed lines show the Q-factor when only linear equalization (LE) is implemented. b) Q-factor achieved after pruning compared to the one achieved after both pruning and quantization, for different levels of sparsity and for a dataset corresponding to the 1~dBm launch power. The blue and red solid lines correspond to the Q-factor achieved by the original model and the one achieved by this model after quantization, respectively.}
\label{fig:test}
\end{figure}
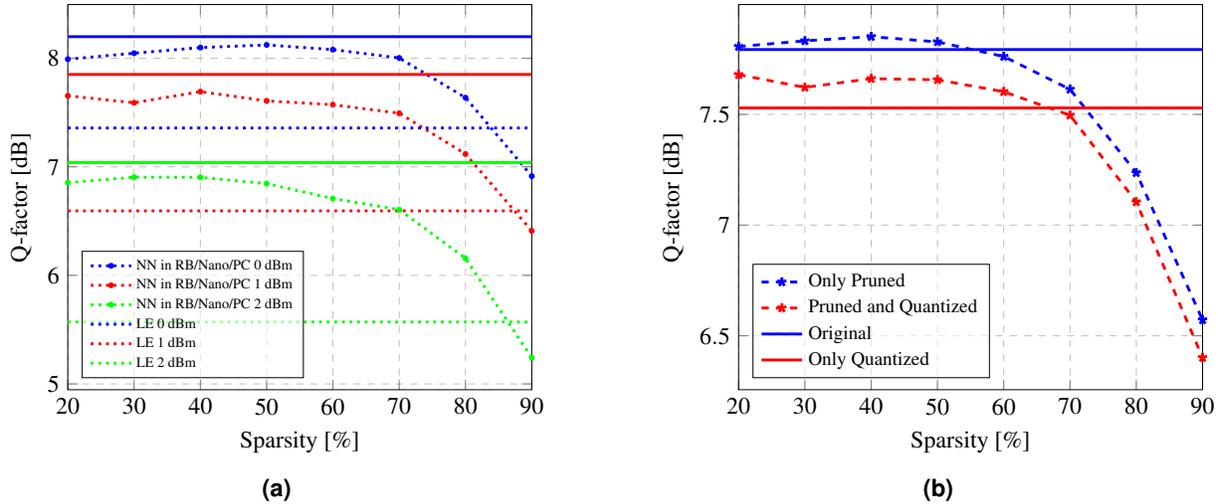

Firstly, we note that the complexity reduction of the equalizer must not affect its performance drastically, i.e. the system's performance is still required to be within an acceptable range.  In Fig.~\ref{fig:figure2}, the Q-factor achieved by the NN equalizer is depicted versus different sparsity values, for the three launch power levels: 0 dBm, blue; 1 dBm, red; and 2 dBm, green. The results are shown using dotted lines and stars, which are those obtained on the PC, Raspberry Pi, and Nvidia Jetson Nano, using the pruned and quantized model. For each of these launch powers, two baselines for the Q-factor are depicted: one corresponds to the level achieved by the uncompressed model, defined by the straight lines, while the other provides the benchmark when we do not employ any NN equalization and use only standard linear chromatic dispersion compensation plus phase/amplitude normalization (LE, linear equalization); the latter levels for the three different launch powers are marked by dotted lines having the appropriate colors.

Fig.~\ref{fig:Qfact_sparsity_vs_quantization} quantifies the impact that each compression technique has on the performance: in that figure, we plotted the Q-factor achieved by the NN equalizer versus different values of sparsity, for the 1 dBm launch power. The blue and red straight lines represent the Q-factor of the original model and the Q-factor achieved by it after being quantized. The dotted lines with asterisks, show the performance of a model that has been only pruned (blue), and the performance in the case of both pruning and quantization (red). It is seen that a substantial reduction of the complexity can be achieved without a dramatic degradation of the performance. The sparsity levels at which the fast deterioration of the performance occurs, are also clearly seen in this figure.

First, it can be observed from Fig.~\ref{fig:figure2} that the quantization and pruning process does not cause a significant performance degradation until a sparsity level equal to 60\% is reached, with just a $4\%$ performance reduction. 
However, when we move to sparsity levels around 90\%, the performance is close to the one achieved using a linear equalization (i.e., the Q-factor curves drop to the levels marked with the dashed lines of the same color). 

We can conclude that when the levels of sparsity are above 60\%, the decrease in the performance is mainly the effect of the quantization process. A nearly 2.5\% drop in the Q-factor value has also been observed when quantizing an already pruned model. Once the levels of sparsity are higher than 60\%, the reduction in performance due to the quantization gets accelerated. Moreover, we observe that some degree of sparsification can even improve the model's performance with respect to the unpruned model. This behavior has already been reported in other studies and it was found that it is specifically pertinent to the over-parametrized models. Thus, the NNs with less complex structures do not show up such an increase in performance due to low-sparsity pruning, making it impossible to achieve such a good performance-complexity ratios~\cite{hoefler2021sparsity, allen2019convergence,neyshabur2018towards,zhu2017prune}.

\subsection*{Computational Complexity Analysis}
\begin{figure}[ht]
    \centering
\begin{tikzpicture}[scale=0.97]
  \begin{axis}[
    ybar,
    symbolic x coords={20,30,40,50,60,70,80,90},
    bar width=0.3cm,
    xtick=data,
    axis y line*=left,
    ymajorgrids = true,
    ylabel = {\textcolor{blue}{Decrease in the size of the model (\%)}},
    xlabel = {Sparsity(\%)},
    ymin=80,ymax=100,
        bar shift={-\pgfplotbarwidth/2},
    ]
    \addplot[style={blue,fill=blue,mark=none}]
           coordinates {(20,82.5049543799699) (30, 83.3463297853847) (40, 84.19614874563543) (50, 85.47509896342947) (60,87.12010211734554) (70, 89.08000019867188) (80,91.16208148527096) (90,93.65491688065283)};
  \end{axis}
  \begin{axis}[
    ybar,
    bar width=0.3cm,
    xtick=\empty,
    axis y line*=right,
    ylabel=axis2,
    ylabel = {\textcolor{red}{Decrease in the complexity of the model (\%)}},
    ymin=65,ymax=100,
        bar shift={\pgfplotbarwidth/2},
    ]
    \addplot[style={red,fill=red,mark=none}]
             coordinates {(20,72.31405413013707) (30,73.82220783515352) (40, 75.33036154016995) (50, 76.83851524518637) (60, 78.3466689502028) (70,  79.85482265521924) (80,81.36297636023568) (90,82.8711300652521)};
\node (A) at (20,69.3) {};
\node (O1) at (10,69.3) {};
\node (O2) at (99,69.3) {};

\draw [color=black, very  thick,dashed] (A -| O1) -- (A -| O2);
 \end{axis}
\end{tikzpicture}
\caption{Complexity and size reduction achieved via pruning and quantization for different levels of sparsity. The dashed black line represents the reference complexity when only quantization is applied.}
\label{fig:figure4}
\end{figure}
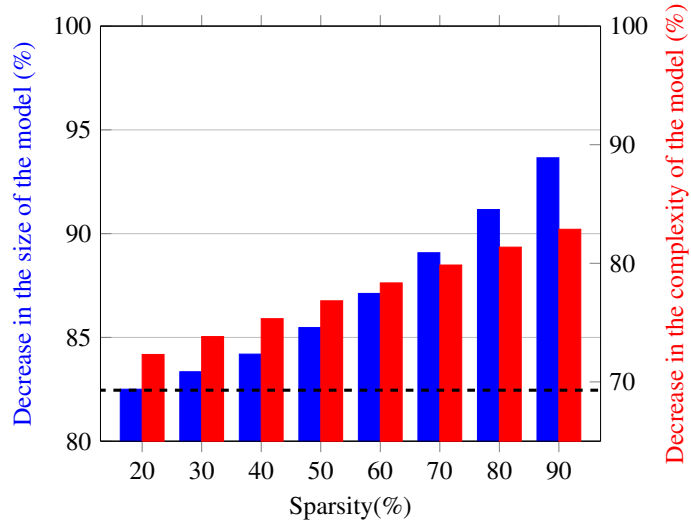

Fig.~\ref{fig:figure4} depicts the reduction in the size of the model as well as the model's computational complexity for different sparsity values, after having applied quantization. For the definition of the metrics used to calculate the computational complexity as well as the size of the models, see the subsections Computational Complexity Metrics and Memory Size Metrics in Methods. 
Overall, we have achieved an 87.12\% reduction in the memory size after pruning 60\% of the NN equalizer weights and quantizing the remaining ones. As a consequence, the size of the model went down from $201.4$ to $25.9$ kilobytes. For the decrease of the model's computational complexity, it goes from $75960427.38$ to $16447962$ bit operations (BoPs) after applying the same compression strategy, which is a $78.34\%$ reduction (see the explicit definition of BoPs in the Methods section). We would like to point out once more that sparsity levels of $60\%$ can be reached without a substantial performance loss. Therefore, approximately the same high level of performance can be achieved with a model that is significantly less complex than the initial NN structure, which is one of the main findings of our work.

It is worth mentioning the individual impact that quantization and pruning have on the computational complexity of the model. When the computational complexity is calculated for a quantized, but unpruned model, the number of BOPs is equal to $23321563$. Therefore, if this value is compared with the already mentioned $75960427$ BoPs for the unpruned and unquantized NN, a reduction in complexity of a 69.3\% is obtained thanks to quantization. As it can be seen in Fig.~\ref{fig:figure4}, the remaining gain comes from the pruning technique, and it grows linearly as indicated in Eq.~(\ref{complfull}).
\subsection*{Online Latency Evaluation}

Numerous deep learning applications are latency-critical, and therefore the inference time must be within the bounds specified by service level objectives. Optical communication applications that employ deep learning techniques are a good example of this. Note that the latency is highly dependent on the NN model implementation and the hardware employed (e.g., FPGA, CPU, GPU). Please refer to the Methods section for more details on the devices' inference time measurements.

When measuring the inference time for the different types of hardware and the quantized model that has had 60\% of its weights pruned, the results are:

\begin{itemize}
    \item Latency Raspberry Pi : $\mu = 0.81~s$ and $\sigma = \pm 0.035 $ 
    \item Latency Nvidia Jetson Nano: $\mu = 0.53~ s$ and $\sigma = \pm 0.022$
    \item Latency PC: $\mu = 0.1~s $ and $\sigma =0.006  $
    
\end{itemize}

In the case of the  unpruned and unquantized model: 

\begin{itemize}
    \item Latency Raspberry Pi : $\mu = 1.84~s$ and $\sigma = \pm 0.08 $ 
    \item Latency Nvidia Jetson Nano: $\mu = 1.22 ~s$ and $\sigma = \pm 0.052 s$
    \item Latency PC: $\mu = 0.18~s $ and $\sigma = \pm 0.008 $
    
\end{itemize}

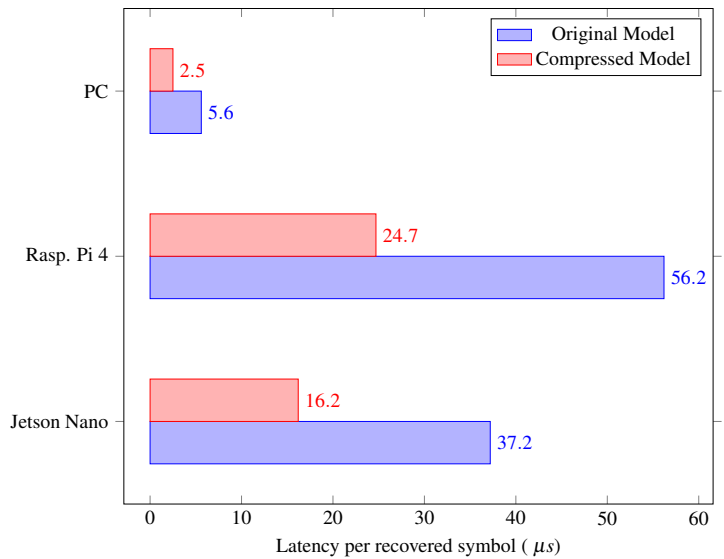
\begin{figure}[htb]
     \centering
     \hspace{-15mm}
     \resizebox{0.55\columnwidth}{!}{
\begin{tikzpicture}
  \centering
  \begin{axis}[
    width=12cm,
    xlabel=Latency per recovered symbol ( $\mu s$),
    xbar=0pt,
    bar width=0.75cm,
    enlarge y limits=0.25,
      ytick=data,
      symbolic y coords={%
        {Jetson Nano},
        {Rasp. Pi 4}, 
        {PC}
      },
      nodes near coords,
      nodes near coords align={horizontal},
    ]

    \addplot +[area legend] coordinates { 
      (5.6,{PC})
      (56.2,{Rasp. Pi 4}) 
      (37.2,{Jetson Nano}) 
};

    \addplot +[area legend] coordinates {
      (2.5,{PC})
      (24.7,{Rasp. Pi 4}) 
      (16.2,{Jetson Nano}) 
    };

    \legend{Original Model, Compressed Model}
  \end{axis}

\end{tikzpicture}
  }
\caption{Summary of the symbol processing (inference) time for compressed NN models (after pruning and quantization) and the original models for three devices under evaluation: Raspberry Pi 4, Jetson Nano, and a standard PC.}
     \label{fig:latencyfinal}
 \end{figure}

Fig.~\ref{fig:latencyfinal} shows the latency of the considered NN model before and after quantization. We notice that the results are expressed in a way that is more appropriate for the task at hand. Thus, latency is defined as the time it takes to process one symbol: we have averaged it over 30k symbols. With the quantized model, we observe approximately a 56\% reduction in latency for all three values of power, when compared to the original model. We must notice that pruning is not taken into account because it does not affect this metric since Tensorflow Lite does not support sparse inference yet, which makes the algorithm still use the same amount of cache memory. Also, we could observe that Raspberry Pi has the longest inference time among our devices. This is in line with the fact that Raspberry is designed as a low-cost and general-purpose single-board computer \cite{hadidi2019characterizing}. On the other hand, the Nvidia Jetson Nano was developed with GPU capabilities, which makes it more suitable for deep learning applications, allowing us to achieve lower latencies. 

\subsection*{Online Energy Consumption Evaluation}
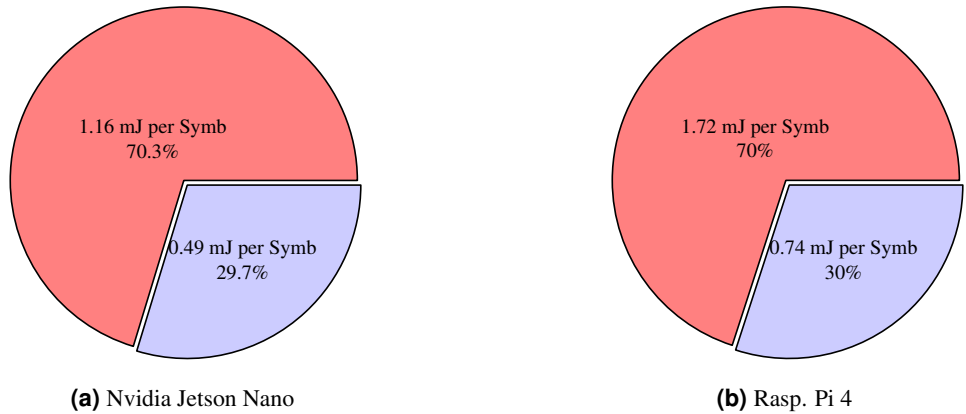
\begin{figure}[ht]
     \centering
       \begin{subfigure}{.45\linewidth}
    \centering\resizebox{0.6\columnwidth}{!}{
\begin{tikzpicture}
    \pie[text=inside,explode={0, 0.1},color={red!50, blue!20}]{70.3/1.16 mJ per Symb, 29.7/ 0.49 mJ per Symb}
\end{tikzpicture}}
    \caption{Nvidia Jetson Nano}
    \label{power_JETSON} 
  \end{subfigure} 
  \begin{subfigure}{.45\linewidth}
    \centering\resizebox{0.6\columnwidth}{!}{
\begin{tikzpicture}
    \pie[text=inside,explode={0, 0.1},color={red!50, blue!20}]{70/1.72 mJ per Symb, 30/0.74 mJ per Symb}
\end{tikzpicture}}

    \caption{Rasp. Pi 4}
    \label{power_RASPI} 
  \end{subfigure}

\caption{Energy consumption for Raspberry Pi 4 and Nvidia Jetson Nano. The blue section represents the energy consumption per recovered symbol when using the compressed model, and its relative energy cost is expressed as a percentage with respect to the sum of the energy consumed by both the original and compressed models. Likewise, the red section describes the energy consumption per recovered symbol when using the original model and its relative energy cost.}
     \label{fig:energyfinal}
 \end{figure}

Within the context of edge computing, not only is speed an important factor, but also power efficiency. In this work, the metric used to evaluate the energy consumption and compare the different types of hardware for the coherent optical channel equalization task is the energy per recovered symbol.
When using a quantized model with a pruning level of 60\%, the average energy consumed during inference for the Raspberry Pi 4 and the Nvidia Jetson Nano is 2.98~W ($\sigma = \pm 0.012 $ ) and 3.03~W ($\sigma = \pm 0.017$), respectively. On the other hand, if the original model is employed, there is an increase in energy consumption of around 3\%, which is congruent with the findings in previous works \cite{sze2017efficient}. Thus, during inference, the Raspberry Pi 4 consumes 3.06~W ($\sigma = \pm 0.011 $ ) and the Nvidia Jetson Nano 3.13~W ($\sigma = \pm 0.015$), respectively.
Multiplying these values by the NN processing times per recovered symbol reported in Fig.~\ref{fig:latencyfinal}, we obtain the results presented in Fig.~\ref{fig:energyfinal}.
We note that Raspberry Pi has the highest energy consumption per recovered symbol. This is a consequence of the lack of a GPU, which results in longer inference times. Thus, Nvidia Jetson Nano consumes 33.78\% less energy than the Raspberry Pi 4. Regarding pruning and quantization, the use of these techniques allows an energy saving of 56.98\% for the Raspberry Pi 4 and a 57.76\% saving for the Nvidia Jetson Nano. 

It must be noticed that although TensorFlow Lite does not support sparse inference and therefore pruning does not help to reduce the inference time, it affects the size of the model. This has a direct effect on the power consumption of the device due to the decrease in the use of resources. In contrast, quantization has a positive effect on both of these parameters thanks to employing lower precision formats and reducing the size of the model. Therefore, it has a stronger effect on energy consumption. This is reflected in the results exposed in this section. Moreover, it is congruent with the findings reported in previous studies \cite{sze2017efficient, yang2017method}.

See the Methods section for more details on the devices' energy consumption measurement.

\section*{Discussion}
In our work, We investigated how we can use pruning and quantization to reduce the complexity of the hardware implementation of a NN-based channel equalizer in a coherent optical transmission system. With this, we tested the implementation of the designed equalizer experimentally, using  Raspberry Pi 4 and the Nvidia Jetson Nano edge devices. It was demonstrated that it is possible to reduce the NN's memory usage by $87.12\%$, and the NN's computational complexity by $78.34\%$ without any serious penalty in performance, thanks to the two aforementioned compression techniques.

Moreover, the effect of using different types of hardware was experimentally characterized by measuring the inference time and energy consumption in both Raspberry Pi 4 and the Nvidia Jetson Nano. We note, however, that we experimented only with the edge devices, and the data from the communication system were obtained via simulations; but we do not expect that the results for the true optical system would seriously differ. It has been demonstrated that Nvidia Jetson Nano allows 34\% faster inference times than Raspberry Pi, and that, thanks to the quantization process, a 56\% inference time reduction can be achieved. Finally, due to the use of pruning and quantization techniques, we achieve 56.98\% energy savings for Raspberry Pi 4 and 57.76\% for the Nvidia Jetson Nano; we also observed that the latter device consumes 33.78\% less energy.

Overall, our findings demonstrate that the usage of pruning and quantization can be a suitable strategy for the implementation of NN-based equalizers that are efficient in high-speed optical transmission systems when deployed on resource-restricted hardware. We believe that these model compression techniques can be used for the deployment of NN-based equalizers in real optical communication systems, and for the development of novel online optical signal processing tools. We hope that our results can also be of interest to the researchers developing sensing and laser systems, where the application of machine learning for field processing and characterization is a rapidly developing area of research\cite{narhi2018machine}.


\section*{Methods}
\subsection*{ Numerical Setup and Neural Network Model}\label{subsec:sw}

We numerically simulated the dual-polarization (DP) transmission of a single-channel signal at 30 GBd. The signal is pre-shaped with a root-raised cosine (RRC) filter with 0.1 roll-off at a sampling rate of 8 samples per symbol. In addition, the signal modulation format is 64-QAM. We considered the case of transmission over 20×50 km links of SMF.  The optical signal propagation along the fiber was simulated by solving the Manakov equation via
split-step Fourier method \cite{AGRAWAL201327} with the resolution of 1 km per step. The considered parameters of the TWC fiber
are: the attenuation parameter $ \alpha = 0.23 dB/km$, the dispersion
coefficient $D = 2.8 ps/(nm \cdot km)$, and the effective nonlinearity coefficient $ \gamma= 2.5 (W \cdot km)^{-1}$. The SSMF parameters are: $\alpha = 0.2 dB/km$, $D = 17 ps/(nm \cdot km)$, and $ \gamma= 1.2 (W \cdot km)^{-1}$. Moreover, after each span, an optical amplifier with the noise figure NF = 4.5 dB was placed to fully compensate fiber
losses and added amplified spontaneous emission (ASE) noise.
At the receiver, a standard Rx-DSP was employed. It consisted of the
full electronic chromatic dispersion compensation (CDC) using
a frequency-domain equalizer, the application of a matched
filter, and the downsampling to the symbol rate. Finally, the
received symbols were normalized (by phase and amplitude)
to the transmitted ones. In this work, no additional transceiver distortions were taken into account.
After the Rx-DSP, the bit error rate (BER) is estimated using the transmitted symbols, received soft symbols, and hard decisions after equalization.

The NN receives as input a tensor with a shape defined by three dimensions: $(B, M, 4)$, where $B$ is the mini-batch size, $M$ is the memory size determined by the number of neighbors $N$ as $M = 2N + 1$, and 4 is the number of features for each symbol, which correspond to the real and imaginary parts of two polarization components. The NN will have to recover the real and imaginary parts of the k-th symbol of one of the polarization. Therefore the shape of the NN output batch can be expressed as $(B, 2)$.
 This task can be treated as a regression or classification one. This aspect has been considered in previous studies and stated that the results achieved by regression and classification algorithms are similar but fewer epochs are needed in the case of regression. Thus, the mean square error (MSE) loss estimator is used in this paper, as it is the standard loss function employed in regression tasks \cite{freire2021neural}.  
The loss function is optimized using the Adam algorithm \cite{kingma2014adam} with the default learning rate equal to 0.001. The maximum number of epochs during the training process was 1000, as it was stopped earlier if the value of the loss function did not change over 150 epochs. After every training epoch, we calculated the BER obtained using the testing dataset.
 The optimal number of neurons and activation functions in each layer of the NN, as well as the memory (input) of the system were inferred employing the Bayesian Optimization algorithm (BO). The values tested for the number of neurons were $n \in [10, 500] $ . For the activation function, the BO had to chose between: "$\tanh$", "ReLu", "sigmoid" and "LeackyReLu". The values tested for the memory (input) of the system were $N \in [5, 50] $ The metric of the BO was the BER, finding the hyperparameters that helped to reduce the BER as much as possible with a validation dataset of $2^{17}$ data points. The final solution was the use of "$\tanh$" as an activation function and 500, 10, and 500 neurons for the first, second, and third layer, respectively. 
 The training and test datasets were composed of independently generated symbols of length $2^{18}$ each. To prevent any possible data periodicity and overestimation \cite{eriksson2017applying, freire2021caveats}, a pseudo-random bit sequence (PRBS) of order 32 was used to generate those datasets with different random seeds for each of them. The periodicity of the data is, therefore, $2^{12}$ times higher than our training dataset size. For the simulation, the Mersenne twister generator \cite{matsumoto1998mersenne} was used with different random seeds. Moreover, the training data was shuffled before being used as an input to the NN. 
 
Finally, we would like to notice an important matter as it is the necessity of the periodical retraining of the equalizer on realistic transmission. In this case, it may be a point of concern. This issue has already been addressed in previous studies \cite{freire2021transfer}, where it has been demonstrated that using transfer learning can drastically reduce the training period and training dataset when changes on the transmission setup occur. 
 
\subsection*{Pruning Technique}
With pruning, the redundant NN elements can be removed to sparsify the network without significantly limiting its ability to carry out a required task \cite{dong2019understanding, hoefler2021sparsity, liang2021pruning}. Thus, networks with a reduced size and computational complexity are obtained, resulting in lower hardware requirements as well as faster prediction times \cite{liang2021pruning, sze2017efficient}. Furthermore, pruning acts as a regularization technique, improving the model quality by helping to reduce overfitting \cite{hoefler2021sparsity}. Moreover, retraining an already pruned NN can help to escape local loss function minima, which can lead to a better prediction accuracy \cite{liang2021pruning}. Thus, less complex models can often be achieved without a noticeable effect on the NN's performance \cite{hoefler2021sparsity}. 

Depending on what is going to be pruned, the sparsification techniques can be classified into two types: model sparsification and ephemeral sparsification \cite{hoefler2021sparsity}. In the first case, the sparsification is permanently applied to the model, while in the second case, the sparsification only takes place during the computing process. In our work, we will use the model sparsification, because of the effects it has on the final NN's computing and memory hardware requirements. Adding to this, the model sparsification can consist in removing not only weights but also larger building blocks, such as neurons, convolutional filters, etc. \cite{hoefler2021sparsity}. Here we apply the specification to just the weights of the network, for the sake of simplicity and as far as it matches the NN structure (the MLP) that is considered.

After having defined what to prune, it is necessary to define when the pruning occurs. Based on this, there are two main types of pruning: static and dynamic \cite{liang2021pruning}. In the static case, the elements are to be removed from the NN after the training, and in this work, to demonstrate the effect, we use the static pruning variant because of its simplicity.

The static pruning is generally carried out in three steps. First, we decide upon what requires to be pruned. A simple approach to define the pruning objects can be to evaluate the NN's performance with and without particular (pruned) elements. However, this poses scalability problems: we have to evaluate the performance when pruning each particular NN's parameters, and there may be millions of these. 

\begin{figure}[ht]
    \centering
    \includegraphics[width=0.55\linewidth]{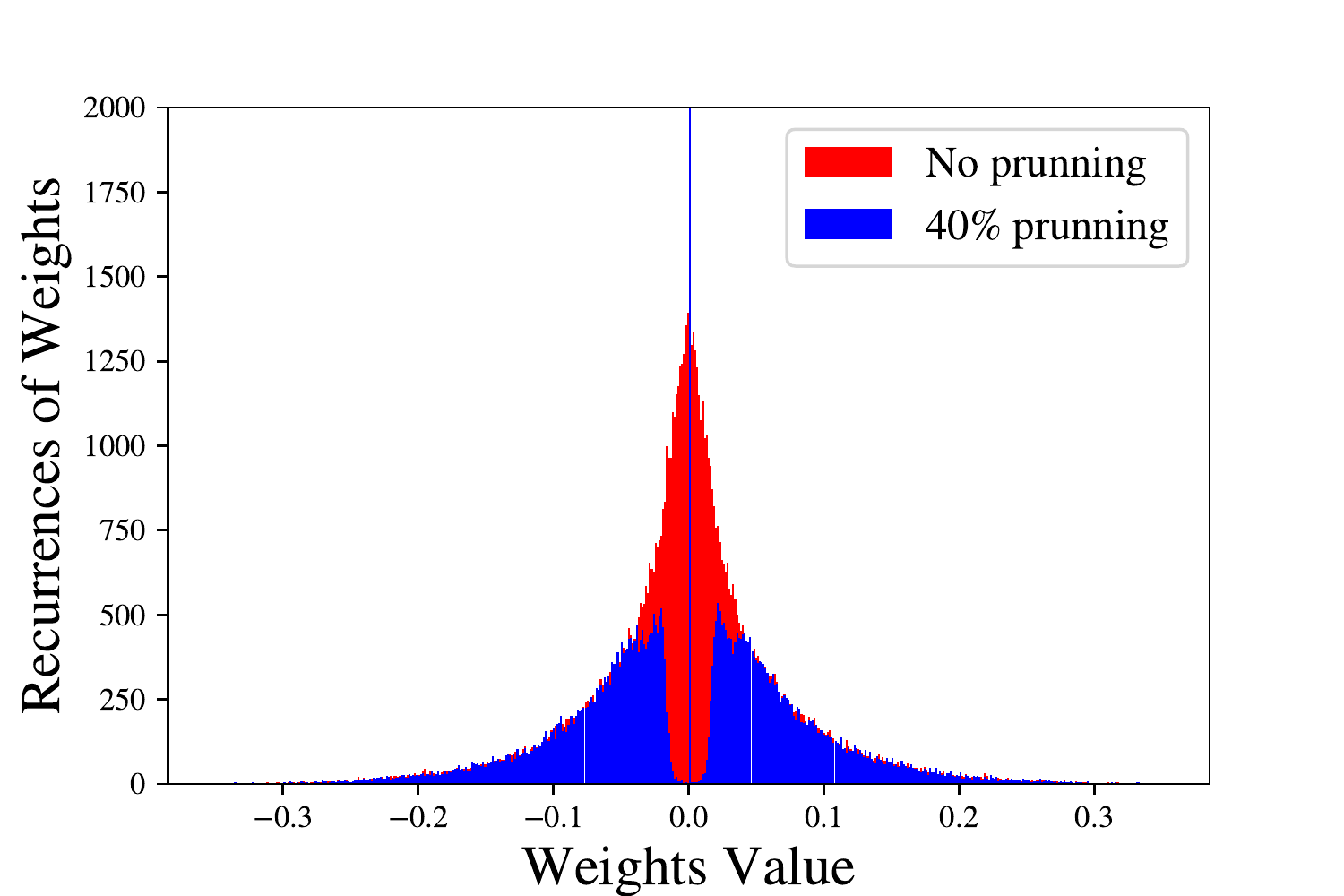}
    \caption{A typical distribution of the weights of the NN-based MLP equalizer without pruning and with pruning when the sparsity level is set to 40\%. }
    \label{fig:prunning_weights}
\end{figure}

Alternatively, it is possible to select the elements to be removed randomly, which can be done faster \cite{hoefler2021sparsity, bondarenko2015neurons, hu2016network}. Following this latter approach, we beforehand decided to prune the weights. Once it has been decided which elements are to be pruned, it is necessary to establish the criteria for how the elements are to be removed from the NN, ensuring that the high levels of sparsity are achieved without a significant loss in performance. When pruning the weights of the NN, it is possible to remove them based on different aspects: considering their magnitude (i.e., the weights having values close to zero are to be pruned, with the pruning percentage is defined by the sparsity level we aim to achieve), or their similarity (if two weights have a similar value, only one of those is kept); we mention that the other selection procedures also exist \cite{hoefler2021sparsity, hu2016network}. Here, we pick the relatively simple weights pruning strategy based on their magnitude. In Fig.~\ref{fig:prunning_weights} we show the impact when we have pruned our NN equalizer by 40\%. When comparing the weight distributions of the original and pruned models, it is clear that the sparsity level defines the number of weights that need to be pruned. Thus, the pruning process starts by removing the smallest weight and continues until the desired sparsity level is reached. Finally, a retraining or fine-tuning phase should be done, to reduce the degradation in the modified NN performance \cite{liang2021pruning}. 

When carrying out pruning using the Tensorflow Model Optimization API, it is necessary to define a pruning Schedule to control this process by notifying at each step the level at which the layer should be pruned \cite{bartoldson2020generalization}. In this work, the schedule known as Polynomial Decay is employed. The main characteristic of this type of schedule is that a polynomial sparsity function is built. In this case, the power of the function is equal to 3 and the pruning takes place every 50 steps. This means that during the last steps higher ratios of sparsification are employed (e.g. more weights are removed), speeding up the pruning process. On the other hand, if the power of the function were negative, pruning would be slowed-down. The model starts with a 0\% sparsity and the process takes place during 300 epochs. This is approximately 35 \% of the number of iterations required for training the original model. It is the objective of future works to optimize the hyperparameters of the pruning process, improve its efficiency and reduce the cost related to a high number of iterations.

 \subsection*{Quantization Technique}

Besides the reduction in the number of operations involved in the NN signal processing, the precision of such arithmetic operations is yet another crucial factor when determining the model's complexity and, therefore, the inference latency, as well as equalizer's memory and energy requirements \cite{choukroun2019low,yang2019quantization,wu2020integer,sze2017efficient}. The process of approximating a continuous variable with a specified set of discrete values is known as quantization. The number of discrete values will determine the number of bits necessary to represent the data. Thus, when applying this technique in the context of deep learning, the objective is decreasing the numeric precision used to encode the weights and activations of the models, avoiding a noticeable decrease in NN's performance \cite{han2016deep, wu2020integer}.

Using low-precision formats allows us to speed up math-intensive operations, such as convolution and matrix multiplication \cite{wu2020integer}. On the other hand, the inference (signal processing) time depends not only on the format representation of the digits involved in the mathematical operations but is also affected by transporting the data from memory to the computing elements \cite{sze2017efficient, yang2017method}. Moreover, heat is generated during the latter process and, therefore, using a lower-precision representation can result in energy savings \cite{sze2017efficient}. Finally, another benefit of using low-precision formats is that a reduced number of bits is needed to store the data, which reduces the memory footprint and size requirements \cite{wu2020integer,sze2017efficient}.

FP32 has been traditionally used as the numerical format for encoding weights and activations (output of the neurons) in a NN, to take advantage of a wider dynamic range. However, as it has already been mentioned, this results in higher inference times, which is an important factor when a real-time signal processing is considered \cite{han2016deep}.
A variety of alternatives to the FP32 numerical format for NN's elements representation have been proposed lately, to reduce the inference time, as well as to decrease the hardware requirements. For example, it is becoming popular to train NNs in FP16 formats, as it is supported by most deep learning accelerators \cite{han2016deep}. On the other hand, math-intensive tensor operations executed on INT8 types can see up to a 16$\times$ speed-up compared to the same operations in FP32. Moreover, memory-limited operations could see up to a 4$\times$ speed-up compared to the FP32 version \cite{hawks2021ps,sze2017efficient, liang2021pruning, wu2020integer}. 
Therefore, in addition to pruning, we will reduce the precision of the weights and activations to further decrease the computational complexity of the equalizer, employing the technique known as integer quantization \cite{wu2020integer}.

The integer quantization maps a floating point value $x\in [\alpha,\,\beta ]$ to a bit integer $x_{q}\in [\alpha_{q},\,\beta_{q} ]$. This mapping can be defined mathematically using the following formula: $x_{q} = \mathrm{round} \left(\frac{1}{s}x + z\right)$, where $s$ (a positive floating point number) is known as the scale, and $z$ is the zero point (an integer). The previous equation can be refactored in order to take into account that if $x$ is outside of the range $[\alpha,\,\beta]$, then $x_{q}$ is outside of $[\alpha_{q}, \, \beta_{q}]$. Thus, it is necessary to clip the values when this happen; as a consequence, the mapping formula becomes: $x_{q} = \mathrm{clip}(\mathrm{round} \left[ \frac{1}{s}x + z \right], \alpha_{q}, \beta_{q})$, where the $\mathrm{clip}$ function takes the values \cite{gholami2021survey,liang2021pruning}: $$\mathrm{clip}(x,l,u)=\begin{cases}
l & \text{ if } x< l , \\ 
x & \text{ if } l\leq x\geq u, \\ 
u & \text{ if } x> u .
\end{cases}$$

Integer quantization can take different forms, depending on the spacing between quantization levels and the symmetry of the clipping range (determined by the value of the zero-point $z$) \cite{gholami2021survey}. For the sake of simplicity, in this work, we used symmetric and uniform integer quantization.

The quantization process can occur after the training or during it. The first case is known as post-training quantization (PTQ) and the second one is the quantization aware training \cite{hawks2021ps,sze2017efficient, liang2021pruning}. In PTQ, a trained model has its weight and activations quantified. After this, a small unlabeled calibration set is used to determine the activations’ dynamic ranges \cite{sze2017efficient, wu2020integer,hubara2021accurate,gholami2021survey}. No retraining is needed, which makes this method very popular because of its simplicity and lower data requirements \cite{gholami2021survey, hubara2021accurate}. Nonetheless, when a trained model is directly quantized, this may perturb the trained parameters, moving the model away from the convergence point reached during the training with a floating-point precision. In other words, we notice that PTQ can have accuracy-related issues \cite{gholami2021survey}. 

In this work, the quantization is carried out after the training stage, i.e., we use the PTQ. The calibration process required to estimate the range, i.e, (min, max) of the activations in the model, is done by running a few inference cycles with a small portion of the test dataset. In our case, it consisted of 100 samples. When using the Tensorflow Lite API, the calibration is carried out automatically, and it is not possible to choose the number of cycles.

\subsection*{Computational Complexity Metrics}\label{subsec:cc}
Finally, it is important to discuss how we can correctly evaluate the computational complexity of such models. In this regard, we quantitatively evaluate the reduction of computation complexity achieved by applying pruning and quantization, calculating the number of bits used during an inference step. The most common operations in a NN are multiply-and-accumulate operations (MACs). These are operations of the form $a = a + w \cdot x$, where three terms are involved: firstly, $x$ corresponds to the input signal of the neuron; secondly, $w$ refers to the weight; and, finally, the accumulate variable $a$ \cite{de2019machine}. Traditionally, the network complexity arithmetic has been measured using the number of MAC operations. However, in terms of the DSP processing, the number of BoPs is a more appropriate metric to describe the computational complexity of the model, as for low-precision networks composed of integer operations, it is not possible to measure the computational complexity using FLOPS \cite{baskin2021uniq, hawks2021ps}. Thus, in this work, we use BoPs to quantify the complexity of the equalizer. It is important to notice that within the context of optical channel non-linear compensation, the complexity of NN-based channel equalizers has been traditionally measured taking into account only  the number of multiplications \cite{freire2020complex, deligiannidis2021performance, freire2021caveats}. Thus, the accumulator contribution there was neglected. However, in this project, we aim to have a more general complexity metric and therefore include it in our calculations.

The BOPs measure was proposed for the first time in \cite{baskin2021uniq}, and defined for a convolutional layer that had been quantized:
  \begin{equation}\label{Bops}
    \mathrm{BoPs} = mnk^{2} (\large b_{a}b_{w} + b_{a} +b_{w} + \log_{2}(nk^{2}) \large).
\end{equation} 
In Eq.~(\ref{Bops}), $b_{w}$ and $b_{a}$ are the weight and activation bit-width, respectively; $n$ is the number of input channels, $m$ is the number of output channels, and $k$ defines the filters size (e.g. $k\times k$ filters) \cite{albawi2017understanding}. Taking into account that a MAC operation takes the form: $a = a + w \cdot x$, it is possible to distinguish two contributions in the equation above: one corresponding to the $nk^{2}\cdot b_{0}$ number of additions, where $b_{0} = b_{a} +b_{w} + \log_{2}(nk^{2})$ (e.g. accumulator width in the MAC operations), and the other corresponds to the number of multiplications, e.g. $nk^{2}(\large b_{a}b_{w})$ \cite{baskin2021uniq}. 
 
Eq. (\ref{Bops}) was further adapted for the case of a dense layer that has been both pruned and quantized \cite{tran2021ps}. Thus, it is applicable to our case, as the MLP consists of a series of dense layers arranged one after the other:
  \begin{equation}\label{Bops_dense}
    \mathrm{BoPs}_{i}  = m_{i}n_{i} \large[(1 - f_{p_{i}})b_{a_{i}}b_{w_{i}} + b_{a_{i}} + b_{w_{i}} + \log_{2}(n_{i}) \large],
    \end{equation}
In Eq. (\ref{Bops_dense}), $n$ and $m$ correspond to the number of inputs and outputs, respectively; $b_{w}$ and $b_{a}$ are the bit widths of the weights and activations. The additional term, $f_{p_{i}}$, is the fraction of pruned layer weights, which allows us to take into account the reduction in multiplication operations because of pruning. This is the reason why it only relates to the term $b_{a}b_{w}$ \cite{tran2021ps}.

Therefore, in our case of the MLP with 3 hidden layers, the total number of BOPs is: 
\begin{equation} \label{BopsFull}
\mathrm{BoPs} = \text{BoPs}_{input} + \sum_{i} \text{BoPs}_{i} + \text{BoPs}_{output},
\end{equation}
where $i\in [1,2,3]$, $\mathrm{BoPs_{input}}$ and $\mathrm{BoPs_{output}}$ correspond to the contributions of the input and ouput layers. Eq. (\ref{BopsFull}) can be written in a less compact way as follows:
\begin{align}
\label{complfull}\nonumber
\mathrm{BoPs_{MLP}} &= (n_i n_1b_i + n_1 n_2 b_a  +   n_2 n_3b_a  +  n_3 n_ob_a)(1  -   f_{p}) b_w +  (n_i n_1) (b_i + b_w )\log_{2}(n_{i})\\ 
&+ (n_1 n_2) (b_a + b_w )\log_{2}(n_{1}) +  (n_2 n_3) (b_a + b_w )\log_{2}(n_{2})  + (n_3n_o) (b_a + b_w )\log_{2}(n_{3}),
\end{align}
where $n_{i}$,  $n_{1}$, $n_{2}$, $n_{3}$ ,and $n_{o}$ are the  number of neurons in the input, first, second, third, and output layers, respectively; $b_{w}$, $b_{a}$, $b_{o}$  and $b_{i}$ are the bit width of the weights, activations, input and output, respectively; $f_{p}$ is the fraction of the weights that have been pruned in a layer, which, in our case, is the same for every layer.

\subsection*{Memory Size Metrics}\label{subsec:memsize}
In this work, the size of the model is defined as the number of bytes that it occupies in memory. Moreover, we notice the direct correlation between the value of this metric and the format used to represent the model. Thus, in contrast to the traditional formats used in Tensorflow (e.g .h5 or HDF5 binary data format and .pb or protobuf), a TensorFlow Lite model, it is represented in a special efficient portable format identified by the \texttt{.tflite} file extension. This provides two main advantages: a reduced model's size and lower inference times. Therefore, the deployment of the NN model on a resource-restricted hardware becomes feasible. As a consequence, it would not make sense to compare the models saved in the traditional Tensorflow format with those that have been pruned and quantized as well as converted into Tensorflow Lite. We were aware of this situation during the realization of the procedure and, thus, to avoid overestimating the benefits of pruning and quantization, the unpruned and unquantized models have also been converted to \texttt{.tflite} format. To better understand the implications that this step has, the size of the original model in \texttt{.h5} format would experiment a 96.22\% size reduction after being converted to \texttt{.tflite} format, quantized and pruned (60\% sparsity). On the other hand, if the original model has already been converted to \texttt{.tflite}, the size reduction is 87.12\%. Of course, based on this, always using \texttt{.tflite} format instead of the other conventional ones seems to be the best strategy. The main reason behind not doing this is that a graph that is in \texttt{.tflite} format can not be trained again, so, it only supports an online inference mode. Nevertheless, a model that is, for example, in \texttt{.h5} format, can be trained offline. Therefore, the \texttt{.tflite} is only intended to be used in the context of edge computing.

\subsection*{Memory and Processor Restricted Hardware}
In many deep learning applications, low power consumption and a reduced inference time are especially desirable. Moreover, the use of graphics processing units (GPU) to attain high performance has some costs-related issues which are far from being ultimately solved\cite{hadidi2019characterizing, valladares2021performance}. Therefore, a small, portable, and low-cost hardware is required to bring the solution to this problem. As a result, single-board computers have become popular, and Raspberry Pi 4 and Nvidia Jetson Nano are among the most used ones\cite{hadidi2019characterizing}. Hence, here we analyse the functioning of our NN-based equalizer using these two aforementioned popular hardware types.

\subsubsection*{Raspberry Pi}
Raspberry Pi is a small single-board computer. It is equipped with a Broadcom Video Core VI (32-bit) GPU, Quad-core ARM CortexA72 64-bit 1.5 GHz CPU, 2 USB 2.0 ports, and 2 USB 3.0 ports; for data storage, it uses a MicroSD card. Moreover, connections are provided through a Gigabit Ethernet / WiFi 802.11ac. It uses an OS known as Raspbian and has no GPU capability as well as no specialized hardware accelerator \cite{hadidi2019characterizing, tang2018experimental}.
 
\subsubsection*{NVIDIA Jetson Nano}
NVIDIA Jetson Nano is a small GPU-based single-board computer that allows the parallel operation of multiple NNs. It has a reduced size (100 mm $\times$ 80 mm $\times$ 29 mm) and is equipped with a Maxwell 128-core GPU, Quad-core ARM A57 64-bit 1.4 GHz CPU. Like in the case of Raspberry Pi, a MicroSD card is used to store the data. Finally, connections are established via Gigabit Ethernet and the OS employed is Linux4Tegra, based on Ubuntu 18.04 \cite{valladares2021performance, hadidi2019characterizing}.


\subsection*{Power Measurement}\label{Sec:PowerConsumption}
In this work, together with the latency and accuracy attributed to each model processing, we also address the issue of the power consumption for the NN equalizers implemented in Nvidia Jetson Nano and the Raspberry Pi 4. 

It is possible to measure the power consumption of both the Nvidia Jetson Nano and the Raspberry Pi in different ways. Regarding Nvidia Jetson Nano, there are three onboard sensors located at the power input, at the GPU, and at the CPU. Thus, the precision of the measurements is limited by these sensors. To read the recordings of these sensors, it is possible to do it automatically using the \texttt{tegrastats} tool, or manually by reading \texttt{.sys} files, a pseudo-file system on Linux. By using both approaches, the information of measurements for the power, voltage, and current can be readily collected \cite{holly2020profiling}. In contrast, Raspberry Pi 4 has no system to easily provide power consumption numbers. Some software-based methods have been developed, as well as some empirical estimations \cite{kaup2014powerpi}. However, it has been demonstrated that most of the aforementioned software methods give just an approximation that may not be used if very precise results are required \cite{kaup2014powerpi}. On the other hand, the second empiric strategy to measure the power consumption on Raspberry Pi is specific for this type of hardware and cannot be used in Nvidia Jetson Nano.

\begin{figure}
\centering
\begin{subfigure}[t]{0.42\textwidth}
   \includegraphics[width=1\linewidth]{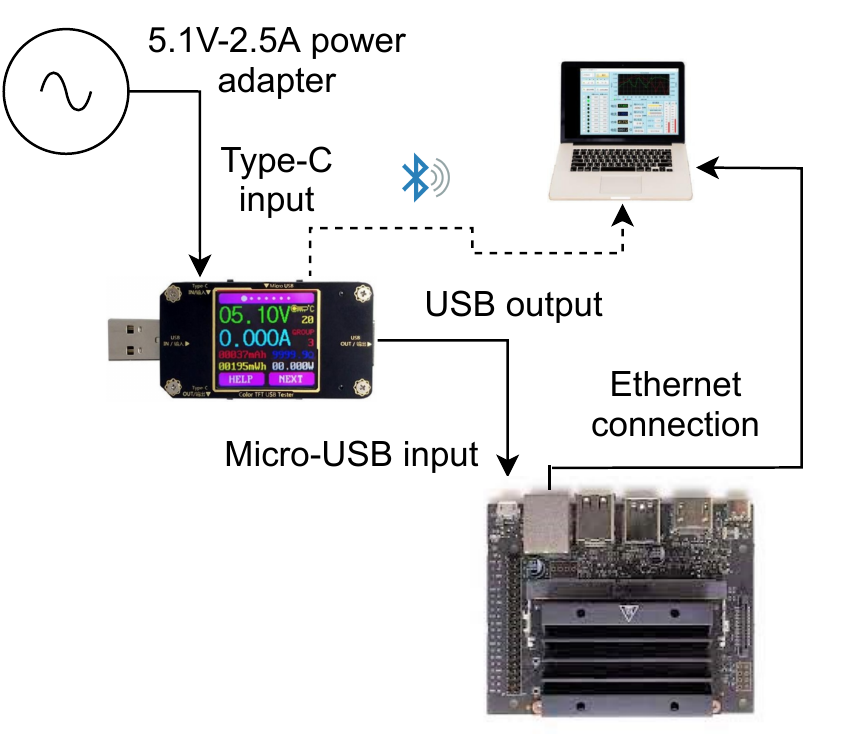}
   \caption{}
   \label{fig:Ng1} 
\end{subfigure}
~
\begin{subfigure}[t]{0.42\textwidth}
   \includegraphics[width=1\linewidth]{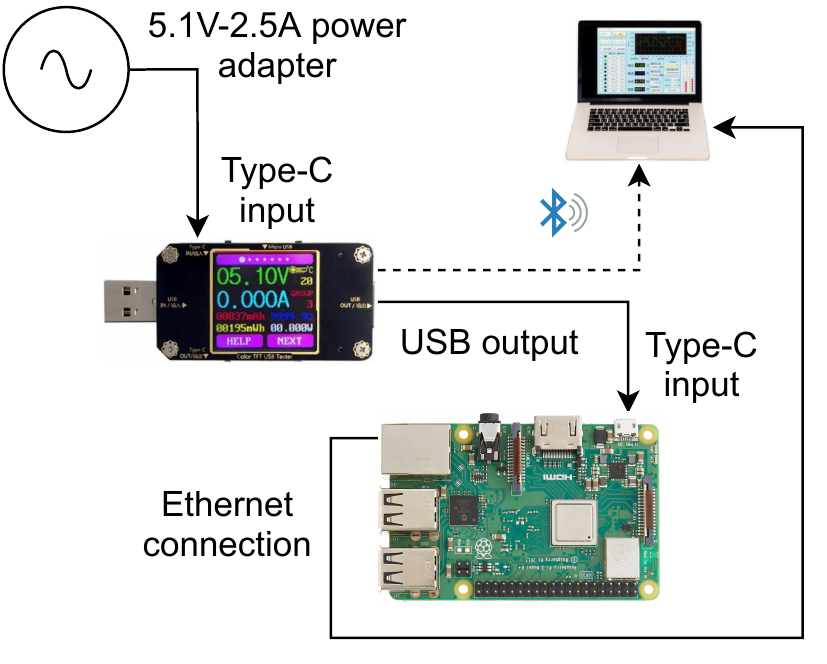}
   \caption{}
   \label{fig:MeasurementetUp}
\end{subfigure}

\caption[Two Power Measurement Set-Ups]{(a) The power measurement set-up for Nivida Jetson Nano, and (b) -- the same for Raspberry Pi.} \label{fig:pw}
\end{figure}

To compare the power consumption of the equalizer on these two types of hardware, it is more accurate and desirable to use the same method in both of them, to avoid any instrumental bias. In this paper, we developed a platform-agnostic method through the use of a digital USB multimeter. The proposed power consumption measurement system addresses the problem of these devices having no onboard shunt resistors; such an approach allows us to easily measure power with an external energy probe. A schematic of the measurement set-ups is given in Fig.~\ref{fig:pw}.

In the case of Raspberry Pi, the power is supplied through a USB type C port via a 5.1~V--2.5~A power adapter. For Nvidia Jetson Nano, the power can be supplied through a Micro-USB connector using a 5.1~V--2.5~A power adapter or a Barrel jack 5~V--4~A (20~W) power supplier. It is possible to change from one configuration to the other by setting a jumper and moving from the 5~W Mode to the 10~W one. To use the same source of power as in Raspberry Pi, the Micro-USB configuration is used. 

As energy is supplied through a USB connection, it is possible to measure the power using a USB digital multimeter. The model used in this work is the A3-B/A3 manufactured by Innovateking-EU. It records voltage, current, impedance, and power consumption. The input voltage and current ranges are 4.5~V--24~V and 0~A--3~A, respectively. Moreover, we can measure the energy in a range that goes from 0~mWh to 99999~mWh. The voltage and current measurement resolution are 0.01~V and 0.001~A, with the measurement accuracies $\pm$0.2\% and $\pm$0.8\%, respectively.

The USB digital multimeter A3-B/A3 comes with the software named UM24C PC Software V1.3, which allows sending the measured data to a computer in real-time, as it is shown in Fig.~\ref{fig:Ng1} and Fig.~\ref{fig:MeasurementetUp} . During the measurement process, no peripherals are connected either to Raspberry Pi or Nvidia Jetson Nano, except for the Ethernet port. This is used for communication over SSH, Fig.~\ref{fig:pw}. Moreover, 25 measures were taken for each device. In each of them, 100 inferences were run, and the power consumption was averaged over them, not taking into account the power consumed during the initialization phase.

\subsection*{Inference Time Measurement} \label{timemeasurement}

To evaluate the inference time for each model, no peripherals are connected either to the Raspberry Pi or to the Nvidia Jetson Nano, except the Ethernet port, which is used to establish communication over the Secure Shell protocol. Moreover, any initialization time (e.g., library loading, data generation, and model weight loading) is ignored because this is a one-time cost that occurs during the device's setup. Furthermore, 25 measures were taken for each device. In each of them, 100 inferences were run (in each inference 30k symbols are recovered) and the inference time was averaged, not taking into account the initialization phase.

\bibliography{sample}

\begin{thebibliography}{10}
\urlstyle{rm}
\expandafter\ifx\csname url\endcsname\relax
  \def\url#1{\texttt{#1}}\fi
\expandafter\ifx\csname urlprefix\endcsname\relax\def\urlprefix{URL }\fi
\expandafter\ifx\csname doiprefix\endcsname\relax\def\doiprefix{DOI: }\fi
\providecommand{\bibinfo}[2]{#2}
\providecommand{\eprint}[2][]{\url{#2}}

\bibitem{Winzer:18}
\bibinfo{author}{Winzer, P.~J.}, \bibinfo{author}{Neilson, D.~T.} \&
  \bibinfo{author}{Chraplyvy, A.~R.}
\newblock \bibinfo{journal}{\bibinfo{title}{Fiber-optic transmission and
  networking: the previous 20 and the next 20 years}}.
\newblock {\emph{\JournalTitle{Opt. Express}}} \textbf{\bibinfo{volume}{26}},
  \bibinfo{pages}{24190--24239}, \doiprefix\url{10.1364/OE.26.024190}
  (\bibinfo{year}{2018}).

\bibitem{Cartledge:17}
\bibinfo{author}{Cartledge, J.~C.}, \bibinfo{author}{Guiomar, F.~P.},
  \bibinfo{author}{Kschischang, F.~R.}, \bibinfo{author}{Liga, G.} \&
  \bibinfo{author}{Yankov, M.~P.}
\newblock \bibinfo{journal}{\bibinfo{title}{Digital signal processing for fiber
  nonlinearities}}.
\newblock {\emph{\JournalTitle{Opt. Express}}} \textbf{\bibinfo{volume}{25}},
  \bibinfo{pages}{1916--1936}, \doiprefix\url{10.1364/OE.25.001916}
  (\bibinfo{year}{2017}).

\bibitem{Rafique}
\bibinfo{author}{Rafique, D.}
\newblock \bibinfo{journal}{\bibinfo{title}{Fiber nonlinearity compensation:
  Commercial applications and complexity analysis}}.
\newblock {\emph{\JournalTitle{Journal of Lightwave Technology}}}
  \textbf{\bibinfo{volume}{34}}, \bibinfo{pages}{544--553},
  \doiprefix\url{10.1109/JLT.2015.2461512} (\bibinfo{year}{2016}).

\bibitem{Dar}
\bibinfo{author}{Dar, R.} \& \bibinfo{author}{Winzer, P.~J.}
\newblock \bibinfo{journal}{\bibinfo{title}{Nonlinear interference mitigation:
  Methods and potential gain}}.
\newblock {\emph{\JournalTitle{Journal of Lightwave Technology}}}
  \textbf{\bibinfo{volume}{35}}, \bibinfo{pages}{903--930},
  \doiprefix\url{10.1109/JLT.2016.2646752} (\bibinfo{year}{2017}).

\bibitem{Zibar02}
\bibinfo{author}{Musumeci, F.} \emph{et~al.}
\newblock \bibinfo{journal}{\bibinfo{title}{An overview on application of
  machine learning techniques in optical networks}}.
\newblock {\emph{\JournalTitle{IEEE Communications Surveys and Tutorials}}}
  \textbf{\bibinfo{volume}{21}}, \bibinfo{pages}{1383 -- 1408},
  \doiprefix\url{10.1109/COMST.2018.2880039} (\bibinfo{year}{2019}).

\bibitem{nevin2021machine}
\bibinfo{author}{Nevin, J.~W.} \emph{et~al.}
\newblock \bibinfo{journal}{\bibinfo{title}{Machine learning for optical fiber
  communication systems: An introduction and overview}}.
\newblock {\emph{\JournalTitle{APL Photonics}}}
  \doiprefix\url{10.1063/5.0070838} (\bibinfo{year}{2021}).

\bibitem{Jar}
\bibinfo{author}{Jarajreh, M.~A.} \emph{et~al.}
\newblock \bibinfo{journal}{\bibinfo{title}{Artificial neural network nonlinear
  equalizer for coherent optical ofdm}}.
\newblock {\emph{\JournalTitle{IEEE Photonics Technology Letters}}}
  \textbf{\bibinfo{volume}{27}}, \bibinfo{pages}{387--390},
  \doiprefix\url{10.1109/LPT.2014.2375960} (\bibinfo{year}{2015}).

\bibitem{hager2018nonlinear}
\bibinfo{author}{H{\"a}ger, C.} \& \bibinfo{author}{Pfister, H.~D.}
\newblock \bibinfo{title}{Nonlinear interference mitigation via deep neural
  networks}.
\newblock In \emph{\bibinfo{booktitle}{2018 Optical Fiber Communications
  Conference and Exposition (OFC)}}, \bibinfo{pages}{1--3}
  (\bibinfo{organization}{IEEE}, \bibinfo{year}{2018}).

\bibitem{zhang2019field}
\bibinfo{author}{Zhang, S.} \emph{et~al.}
\newblock \bibinfo{journal}{\bibinfo{title}{Field and lab experimental
  demonstration of nonlinear impairment compensation using neural networks}}.
\newblock {\emph{\JournalTitle{Nature Communications}}}
  \textbf{\bibinfo{volume}{10}}, \bibinfo{pages}{3033},
  \doiprefix\url{10.1038/s41467-019-10911-9} (\bibinfo{year}{2019}).

\bibitem{freire2021performance}
\bibinfo{author}{Freire, P.~J.} \emph{et~al.}
\newblock \bibinfo{journal}{\bibinfo{title}{Performance versus complexity study
  of neural network equalizers in coherent optical systems}}.
\newblock {\emph{\JournalTitle{Journal of Lightwave Technology}}}
  \textbf{\bibinfo{volume}{39}}, \bibinfo{pages}{6085--6096},
  \doiprefix\url{10.1109/JLT.2021.3096286} (\bibinfo{year}{2021}).

\bibitem{del2020}
\bibinfo{author}{Deligiannidis, S.}, \bibinfo{author}{Bogris, A.},
  \bibinfo{author}{Mesaritakis, C.} \& \bibinfo{author}{Kopsinis, Y.}
\newblock \bibinfo{journal}{\bibinfo{title}{Compensation of fiber
  nonlinearities in digital coherent systems leveraging long short-term memory
  neural networks}}.
\newblock {\emph{\JournalTitle{Journal of Lightwave Technology}}}
  \textbf{\bibinfo{volume}{38}}, \bibinfo{pages}{5991--5999},
  \doiprefix\url{10.1109/JLT.2020.3007919} (\bibinfo{year}{2020}).

\bibitem{deligiannidis2021performance}
\bibinfo{author}{Deligiannidis, S.}, \bibinfo{author}{Mesaritakis, C.} \&
  \bibinfo{author}{Bogris, A.}
\newblock \bibinfo{journal}{\bibinfo{title}{Performance and complexity analysis
  of bi-directional recurrent neural network models versus volterra nonlinear
  equalizers in digital coherent systems}}.
\newblock {\emph{\JournalTitle{Journal of Lightwave Technology}}}
  \textbf{\bibinfo{volume}{39}}, \bibinfo{pages}{5791--5798},
  \doiprefix\url{10.1109/JLT.2021.3092415} (\bibinfo{year}{2021}).

\bibitem{PedroOFC}
\bibinfo{author}{Freire, P.~J.} \emph{et~al.}
\newblock \bibinfo{title}{Experimental study of deep neural network equalizers
  performance in optical links}.
\newblock In \emph{\bibinfo{booktitle}{2021 Optical Fiber Communications
  Conference and Exhibition (OFC)}}, \bibinfo{pages}{1--3}
  (\bibinfo{year}{2021}).

\bibitem{sidelnikov2018}
\bibinfo{author}{Sidelnikov, O.}, \bibinfo{author}{Redyuk, A.} \&
  \bibinfo{author}{Sygletos, S.}
\newblock \bibinfo{journal}{\bibinfo{title}{Equalization performance and
  complexity analysis of dynamic deep neural networks in long haul transmission
  systems}}.
\newblock {\emph{\JournalTitle{Opt. Express}}} \textbf{\bibinfo{volume}{26}},
  \bibinfo{pages}{32765--32776}, \doiprefix\url{10.1364/OE.26.032765}
  (\bibinfo{year}{2018}).

\bibitem{sidelnikov2019methods}
\bibinfo{author}{Sidelnikov, O.~S.}, \bibinfo{author}{Redyuk, A.~A.},
  \bibinfo{author}{Sygletos, S.} \& \bibinfo{author}{Fedoruk, M.~P.}
\newblock \bibinfo{journal}{\bibinfo{title}{Methods for compensation of
  nonlinear effects in multichannel data transfer systems based on dynamic
  neural networks}}.
\newblock {\emph{\JournalTitle{Quantum Electronics}}}
  \textbf{\bibinfo{volume}{49}}, \bibinfo{pages}{1154},
  \doiprefix\url{10.1070/QEL17158} (\bibinfo{year}{2019}).

\bibitem{barry2012digital}
\bibinfo{author}{Barry, J.~R.}, \bibinfo{author}{Lee, E.~A.} \&
  \bibinfo{author}{Messerschmitt, D.~G.}
\newblock \emph{\bibinfo{title}{Digital communication}}
  (\bibinfo{publisher}{New York: Springer}, \bibinfo{year}{3rd edn, 2004}).

\bibitem{ming2021ultralow}
\bibinfo{author}{Ming, H.} \emph{et~al.}
\newblock \bibinfo{journal}{\bibinfo{title}{Ultralow complexity long short-term
  memory network for fiber nonlinearity mitigation in coherent optical
  communication systems}}.
\newblock {\emph{\JournalTitle{arXiv preprint arXiv:2108.10212}}}
  (\bibinfo{year}{2021}).

\bibitem{Kaneda:20}
\bibinfo{author}{Kaneda, N.} \emph{et~al.}
\newblock \bibinfo{title}{Fpga implementation of deep neural network based
  equalizers for high-speed pon}.
\newblock In \emph{\bibinfo{booktitle}{Optical Fiber Communication Conference
  (OFC) 2020}}, \bibinfo{pages}{T4D.2}, \doiprefix\url{10.1364/OFC.2020.T4D.2}
  (\bibinfo{publisher}{Optical Society of America}, \bibinfo{year}{2020}).

\bibitem{blalock2020state}
\bibinfo{author}{Blalock, D.}, \bibinfo{author}{Ortiz, J. J.~G.},
  \bibinfo{author}{Frankle, J.} \& \bibinfo{author}{Guttag, J.}
\newblock \bibinfo{title}{What is the state of neural network pruning?}
  (\bibinfo{year}{2020}).
\newblock \eprint{2003.03033}.

\bibitem{han2016deep}
\bibinfo{author}{Han, S.}, \bibinfo{author}{Mao, H.} \& \bibinfo{author}{Dally,
  W.~J.}
\newblock \bibinfo{title}{Deep compression: Compressing deep neural networks
  with pruning, trained quantization and huffman coding}
  (\bibinfo{year}{2016}).
\newblock \eprint{1510.00149}.

\bibitem{srinivas2017training}
\bibinfo{author}{Srinivas, S.}, \bibinfo{author}{Subramanya, A.} \&
  \bibinfo{author}{Babu, R.~V.}
\newblock \bibinfo{journal}{\bibinfo{title}{Training sparse neural networks}}.
\newblock {\emph{\JournalTitle{2017 IEEE Conference on Computer Vision and
  Pattern Recognition Workshops (CVPRW)}}} \bibinfo{pages}{455--462}
  (\bibinfo{year}{2017}).

\bibitem{hawks2021ps}
\bibinfo{author}{Hawks, B.} \emph{et~al.}
\newblock \bibinfo{journal}{\bibinfo{title}{Ps and qs: Quantization-aware
  pruning for efficient low latency neural network inference}}.
\newblock {\emph{\JournalTitle{Frontiers in Artificial Intelligence}}}
  \textbf{\bibinfo{volume}{4}}, \doiprefix\url{10.3389/frai.2021.676564}
  (\bibinfo{year}{2021}).

\bibitem{sze2017efficient}
\bibinfo{author}{Sze, V.}, \bibinfo{author}{Chen, Y.-H.},
  \bibinfo{author}{Yang, T.-J.} \& \bibinfo{author}{Emer, J.~S.}
\newblock \bibinfo{journal}{\bibinfo{title}{Efficient processing of deep neural
  networks: A tutorial and survey}}.
\newblock {\emph{\JournalTitle{Proceedings of the IEEE}}}
  \textbf{\bibinfo{volume}{105}}, \bibinfo{pages}{2295--2329},
  \doiprefix\url{10.1109/JPROC.2017.2761740} (\bibinfo{year}{2017}).

\bibitem{liang2021pruning}
\bibinfo{author}{Liang, T.}, \bibinfo{author}{Glossner, J.},
  \bibinfo{author}{Wang, L.}, \bibinfo{author}{Shi, S.} \&
  \bibinfo{author}{Zhang, X.}
\newblock \bibinfo{title}{Pruning and quantization for deep neural network
  acceleration: A survey} (\bibinfo{year}{2021}).
\newblock \eprint{2101.09671}.

\bibitem{fujisawa2021}
\bibinfo{author}{Fujisawa, S.} \emph{et~al.}
\newblock \bibinfo{journal}{\bibinfo{title}{Weight pruning techniques towards
  photonic implementation of nonlinear impairment compensation using neural
  networks}}.
\newblock {\emph{\JournalTitle{Journal of Lightwave Technology}}}
  \doiprefix\url{10.1109/JLT.2021.3117609} (\bibinfo{year}{2021}).

\bibitem{li2021high}
\bibinfo{author}{Li, M.}, \bibinfo{author}{Zhang, W.}, \bibinfo{author}{Chen,
  Q.} \& \bibinfo{author}{He, Z.}
\newblock \bibinfo{journal}{\bibinfo{title}{High-throughput hardware deployment
  of pruned neural network based nonlinear equalization for 100-gbps
  short-reach optical interconnect}}.
\newblock {\emph{\JournalTitle{Optics Letters}}} \textbf{\bibinfo{volume}{46}},
  \bibinfo{pages}{4980--4983} (\bibinfo{year}{2021}).

\bibitem{oliari2020revisiting}
\bibinfo{author}{Oliari, V.} \emph{et~al.}
\newblock \bibinfo{journal}{\bibinfo{title}{Revisiting efficient multi-step
  nonlinearity compensation with machine learning: An experimental
  demonstration}}.
\newblock {\emph{\JournalTitle{Journal of Lightwave Technology}}}
  \textbf{\bibinfo{volume}{38}}, \bibinfo{pages}{3114--3124}
  (\bibinfo{year}{2020}).

\bibitem{Koike2021}
\bibinfo{author}{Koike-Akino, T.}, \bibinfo{author}{Wang, Y.},
  \bibinfo{author}{Kojima, K.}, \bibinfo{author}{Parsons, K.} \&
  \bibinfo{author}{Yoshida, T.}
\newblock \bibinfo{title}{Zero-multiplier sparse dnn equalization for
  fiber-optic qam systems with probabilistic amplitude shaping}.
\newblock In \emph{\bibinfo{booktitle}{2021 European Conference on Optical
  Communications (ECOC)}}, \bibinfo{pages}{1--4} (\bibinfo{organization}{IEEE},
  \bibinfo{year}{2021}).

\bibitem{freire2021transfer}
\bibinfo{author}{Freire, P.~J.} \emph{et~al.}
\newblock \bibinfo{journal}{\bibinfo{title}{Transfer learning for neural
  networks-based equalizers in coherent optical systems}}.
\newblock {\emph{\JournalTitle{J. Lightwave Technol.}}}
  \textbf{\bibinfo{volume}{39}}, \bibinfo{pages}{6733--6745},
  \doiprefix\url{10.1109/JLT.2021.3108006} (\bibinfo{year}{2021}).

\bibitem{pelikan1999boa}
\bibinfo{author}{Pelikan, M.}, \bibinfo{author}{Goldberg, D.~E.},
  \bibinfo{author}{Cant{\'u}-Paz, E.} \emph{et~al.}
\newblock \bibinfo{title}{Boa: The bayesian optimization algorithm}.
\newblock In \emph{\bibinfo{booktitle}{Proceedings of the genetic and
  evolutionary computation conference GECCO-99}}, vol.~\bibinfo{volume}{1},
  \bibinfo{pages}{525--532} (\bibinfo{organization}{Citeseer},
  \bibinfo{year}{1999}).

\bibitem{tensorflow2015-whitepaper}
\bibinfo{author}{Abadi, M.} \emph{et~al.}
\newblock \bibinfo{title}{{TensorFlow}: Large-scale machine learning on
  heterogeneous systems} (\bibinfo{year}{2015}).
\newblock \bibinfo{note}{Software available from tensorflow.org}.

\bibitem{hoefler2021sparsity}
\bibinfo{author}{Hoefler, T.}, \bibinfo{author}{Alistarh, D.},
  \bibinfo{author}{Ben-Nun, T.}, \bibinfo{author}{Dryden, N.} \&
  \bibinfo{author}{Peste, A.}
\newblock \bibinfo{title}{Sparsity in deep learning: Pruning and growth for
  efficient inference and training in neural networks} (\bibinfo{year}{2021}).
\newblock \eprint{2102.00554}.

\bibitem{allen2019convergence}
\bibinfo{author}{Allen-Zhu, Z.}, \bibinfo{author}{Li, Y.} \&
  \bibinfo{author}{Song, Z.}
\newblock \bibinfo{title}{A convergence theory for deep learning via
  over-parameterization}.
\newblock In \emph{\bibinfo{booktitle}{International Conference on Machine
  Learning}}, \bibinfo{pages}{242--252} (\bibinfo{organization}{PMLR},
  \bibinfo{year}{2019}).

\bibitem{neill2020overview}
\bibinfo{author}{Neill, J.~O.}
\newblock \bibinfo{title}{An overview of neural network compression}
  (\bibinfo{year}{2020}).
\newblock \eprint{2006.03669}.

\bibitem{neyshabur2018towards}
\bibinfo{author}{Neyshabur, B.}, \bibinfo{author}{Li, Z.},
  \bibinfo{author}{Bhojanapalli, S.}, \bibinfo{author}{LeCun, Y.} \&
  \bibinfo{author}{Srebro, N.}
\newblock \bibinfo{journal}{\bibinfo{title}{Towards understanding the role of
  over-parametrization in generalization of neural networks}}.
\newblock {\emph{\JournalTitle{arXiv preprint arXiv:1805.12076}}}
  (\bibinfo{year}{2018}).

\bibitem{zhu2017prune}
\bibinfo{author}{Zhu, M.} \& \bibinfo{author}{Gupta, S.}
\newblock \bibinfo{journal}{\bibinfo{title}{To prune, or not to prune:
  exploring the efficacy of pruning for model compression}}.
\newblock {\emph{\JournalTitle{arXiv preprint arXiv:1710.01878}}}
  (\bibinfo{year}{2017}).

\bibitem{hadidi2019characterizing}
\bibinfo{author}{Hadidi, R.} \emph{et~al.}
\newblock \bibinfo{title}{Characterizing the deployment of deep neural networks
  on commercial edge devices}.
\newblock In \emph{\bibinfo{booktitle}{2019 IEEE International Symposium on
  Workload Characterization (IISWC)}}, \bibinfo{pages}{35--48}
  (\bibinfo{organization}{IEEE}, \bibinfo{year}{2019}).

\bibitem{yang2017method}
\bibinfo{author}{Yang, T.-J.}, \bibinfo{author}{Chen, Y.-H.},
  \bibinfo{author}{Emer, J.} \& \bibinfo{author}{Sze, V.}
\newblock \bibinfo{title}{A method to estimate the energy consumption of deep
  neural networks}.
\newblock In \emph{\bibinfo{booktitle}{2017 51st asilomar conference on
  signals, systems, and computers}}, \bibinfo{pages}{1916--1920}
  (\bibinfo{organization}{IEEE}, \bibinfo{year}{2017}).

\bibitem{narhi2018machine}
\bibinfo{author}{N{\"a}rhi, M.} \emph{et~al.}
\newblock \bibinfo{journal}{\bibinfo{title}{Machine learning analysis of
  extreme events in optical fibre modulation instability}}.
\newblock {\emph{\JournalTitle{Nature Communications}}}
  \textbf{\bibinfo{volume}{9}}, \bibinfo{pages}{4923},
  \doiprefix\url{10.1038/s41467-018-07355-y} (\bibinfo{year}{2018}).

\bibitem{AGRAWAL201327}
\bibinfo{author}{Agrawal, G.}
\newblock \bibinfo{title}{Chapter 2 - pulse propagation in fibers}.
\newblock In \bibinfo{editor}{Agrawal, G.} (ed.)
  \emph{\bibinfo{booktitle}{Nonlinear Fiber Optics (Fifth Edition)}}, Optics
  and Photonics, \bibinfo{pages}{27--56},
  \doiprefix\url{https://doi.org/10.1016/B978-0-12-397023-7.00002-4}
  (\bibinfo{publisher}{Academic Press}, \bibinfo{address}{Boston},
  \bibinfo{year}{2013}), \bibinfo{edition}{fifth edition} edn.

\bibitem{freire2021neural}
\bibinfo{author}{Freire, P.~J.}, \bibinfo{author}{Prilepsky, J.~E.},
  \bibinfo{author}{Osadchuk, Y.}, \bibinfo{author}{Turitsyn, S.~K.} \&
  \bibinfo{author}{Aref, V.}
\newblock \bibinfo{journal}{\bibinfo{title}{Neural networks based
  post-equalization in coherent optical systems: regression versus
  classification}}.
\newblock {\emph{\JournalTitle{arXiv preprint arXiv:2109.13843}}}
  (\bibinfo{year}{2021}).

\bibitem{kingma2014adam}
\bibinfo{author}{Kingma, D.~P.} \& \bibinfo{author}{Ba, J.}
\newblock \bibinfo{journal}{\bibinfo{title}{Adam: A method for stochastic
  optimization}}.
\newblock {\emph{\JournalTitle{arXiv preprint arXiv:1412.6980}}}
  (\bibinfo{year}{2014}).

\bibitem{eriksson2017applying}
\bibinfo{author}{Eriksson, T.~A.}, \bibinfo{author}{B{\"u}low, H.} \&
  \bibinfo{author}{Leven, A.}
\newblock \bibinfo{journal}{\bibinfo{title}{Applying neural networks in optical
  communication systems: Possible pitfalls}}.
\newblock {\emph{\JournalTitle{IEEE Photonics Technology Letters}}}
  \textbf{\bibinfo{volume}{29}}, \bibinfo{pages}{2091--2094}
  (\bibinfo{year}{2017}).

\bibitem{freire2021caveats}
\bibinfo{author}{Freire, P.~J.} \emph{et~al.}
\newblock \bibinfo{journal}{\bibinfo{title}{Neural networks-based equalizers
  for coherent optical transmission: Caveats and pitfalls}}.
\newblock {\emph{\JournalTitle{arXiv preprint arXiv:2109.14942}}}
  (\bibinfo{year}{2021}).

\bibitem{matsumoto1998mersenne}
\bibinfo{author}{Matsumoto, M.} \& \bibinfo{author}{Nishimura, T.}
\newblock \bibinfo{journal}{\bibinfo{title}{Mersenne twister: a
  623-dimensionally equidistributed uniform pseudo-random number generator}}.
\newblock {\emph{\JournalTitle{ACM Transactions on Modeling and Computer
  Simulation (TOMACS)}}} \textbf{\bibinfo{volume}{8}}, \bibinfo{pages}{3--30}
  (\bibinfo{year}{1998}).

\bibitem{dong2019understanding}
\bibinfo{author}{Dong, X.} \& \bibinfo{author}{Zhou, L.}
\newblock \bibinfo{title}{Understanding over-parameterized deep networks by
  geometrization} (\bibinfo{year}{2019}).
\newblock \eprint{1902.03793}.

\bibitem{bondarenko2015neurons}
\bibinfo{author}{Bondarenko, A.}, \bibinfo{author}{Borisov, A.} \&
  \bibinfo{author}{Alekseeva, L.}
\newblock \bibinfo{title}{Neurons vs weights pruning in artificial neural
  networks}.
\newblock In \emph{\bibinfo{booktitle}{ENVIRONMENT. TECHNOLOGIES. RESOURCES.
  Proceedings of the International Scientific and Practical Conference}},
  vol.~\bibinfo{volume}{3}, \bibinfo{pages}{22--28} (\bibinfo{year}{2015}).

\bibitem{hu2016network}
\bibinfo{author}{Hu, H.}, \bibinfo{author}{Peng, R.}, \bibinfo{author}{Tai, Y.}
  \& \bibinfo{author}{Tang, C.}
\newblock \bibinfo{journal}{\bibinfo{title}{Network trimming: {A} data-driven
  neuron pruning approach towards efficient deep architectures}}.
\newblock {\emph{\JournalTitle{CoRR}}}
  \textbf{\bibinfo{volume}{abs/1607.03250}} (\bibinfo{year}{2016}).
\newblock \eprint{1607.03250}.

\bibitem{bartoldson2020generalization}
\bibinfo{author}{Bartoldson, B.}, \bibinfo{author}{Morcos, A.},
  \bibinfo{author}{Barbu, A.} \& \bibinfo{author}{Erlebacher, G.}
\newblock \bibinfo{journal}{\bibinfo{title}{The generalization-stability
  tradeoff in neural network pruning}}.
\newblock {\emph{\JournalTitle{Advances in Neural Information Processing
  Systems}}} \textbf{\bibinfo{volume}{33}}, \bibinfo{pages}{20852--20864}
  (\bibinfo{year}{2020}).

\bibitem{choukroun2019low}
\bibinfo{author}{Choukroun, Y.}, \bibinfo{author}{Kravchik, E.},
  \bibinfo{author}{Yang, F.} \& \bibinfo{author}{Kisilev, P.}
\newblock \bibinfo{title}{Low-bit quantization of neural networks for efficient
  inference} (\bibinfo{year}{2019}).
\newblock \eprint{1902.06822}.

\bibitem{yang2019quantization}
\bibinfo{author}{Yang, J.} \emph{et~al.}
\newblock \bibinfo{title}{Quantization networks} (\bibinfo{year}{2019}).
\newblock \eprint{1911.09464}.

\bibitem{wu2020integer}
\bibinfo{author}{Wu, H.}, \bibinfo{author}{Judd, P.}, \bibinfo{author}{Zhang,
  X.}, \bibinfo{author}{Isaev, M.} \& \bibinfo{author}{Micikevicius, P.}
\newblock \bibinfo{journal}{\bibinfo{title}{Integer quantization for deep
  learning inference: Principles and empirical evaluation}}.
\newblock {\emph{\JournalTitle{arXiv preprint arXiv:2004.09602}}}
  (\bibinfo{year}{2020}).

\bibitem{gholami2021survey}
\bibinfo{author}{Gholami, A.} \emph{et~al.}
\newblock \bibinfo{journal}{\bibinfo{title}{A survey of quantization methods
  for efficient neural network inference}}.
\newblock {\emph{\JournalTitle{arXiv preprint arXiv:2103.13630}}}
  (\bibinfo{year}{2021}).

\bibitem{hubara2021accurate}
\bibinfo{author}{Hubara, I.}, \bibinfo{author}{Nahshan, Y.},
  \bibinfo{author}{Hanani, Y.}, \bibinfo{author}{Banner, R.} \&
  \bibinfo{author}{Soudry, D.}
\newblock \bibinfo{title}{Accurate post training quantization with small
  calibration sets}.
\newblock In \emph{\bibinfo{booktitle}{International Conference on Machine
  Learning}}, \bibinfo{pages}{4466--4475} (\bibinfo{organization}{PMLR},
  \bibinfo{year}{2021}).

\bibitem{de2019machine}
\bibinfo{author}{de~Lima, T.~F.} \emph{et~al.}
\newblock \bibinfo{journal}{\bibinfo{title}{Machine learning with neuromorphic
  photonics}}.
\newblock {\emph{\JournalTitle{J. Lightwave Technol.}}}
  \textbf{\bibinfo{volume}{37}}, \bibinfo{pages}{1515--1534}
  (\bibinfo{year}{2019}).

\bibitem{baskin2021uniq}
\bibinfo{author}{Baskin, C.} \emph{et~al.}
\newblock \bibinfo{journal}{\bibinfo{title}{Uniq: Uniform noise injection for
  non-uniform quantization of neural networks}}.
\newblock {\emph{\JournalTitle{ACM Transactions on Computer Systems}}}
  \textbf{\bibinfo{volume}{37}}, \doiprefix\url{10.1145/3444943}
  (\bibinfo{year}{2021}).

\bibitem{freire2020complex}
\bibinfo{author}{Freire, P.~J.} \emph{et~al.}
\newblock \bibinfo{journal}{\bibinfo{title}{Complex-valued neural network
  design for mitigation of signal distortions in optical links}}.
\newblock {\emph{\JournalTitle{Journal of Lightwave Technology}}}
  \textbf{\bibinfo{volume}{39}}, \bibinfo{pages}{1696--1705},
  \doiprefix\url{10.1109/JLT.2020.3042414} (\bibinfo{year}{2021}).

\bibitem{albawi2017understanding}
\bibinfo{author}{Albawi, S.}, \bibinfo{author}{Mohammed, T.~A.} \&
  \bibinfo{author}{Al-Zawi, S.}
\newblock \bibinfo{title}{Understanding of a convolutional neural network}.
\newblock In \emph{\bibinfo{booktitle}{2017 international conference on
  engineering and technology (ICET)}}, \bibinfo{pages}{1--6}
  (\bibinfo{organization}{Ieee}, \bibinfo{year}{2017}).

\bibitem{tran2021ps}
\bibinfo{author}{Tran, N.} \emph{et~al.}
\newblock \bibinfo{journal}{\bibinfo{title}{Ps and qs: Quantization-aware
  pruning for efficient low latency neural network inference}}.
\newblock {\emph{\JournalTitle{Frontiers in Artificial Intelligence}}}
  \textbf{\bibinfo{volume}{4}}, \bibinfo{pages}{94} (\bibinfo{year}{2021}).

\bibitem{valladares2021performance}
\bibinfo{author}{Valladares, S.}, \bibinfo{author}{Toscano, M.},
  \bibinfo{author}{Tufi{\~n}o, R.}, \bibinfo{author}{Morillo, P.} \&
  \bibinfo{author}{Vallejo-Huanga, D.}
\newblock \bibinfo{title}{Performance evaluation of the nvidia jetson nano
  through a real-time machine learning application}.
\newblock In \emph{\bibinfo{booktitle}{International Conference on Intelligent
  Human Systems Integration}}, \bibinfo{pages}{343--349}
  (\bibinfo{organization}{Springer}, \bibinfo{year}{2021}).

\bibitem{tang2018experimental}
\bibinfo{author}{Tang, R.}, \bibinfo{author}{Wang, W.}, \bibinfo{author}{Tu,
  Z.} \& \bibinfo{author}{Lin, J.}
\newblock \bibinfo{title}{An experimental analysis of the power consumption of
  convolutional neural networks for keyword spotting}.
\newblock In \emph{\bibinfo{booktitle}{2018 IEEE International Conference on
  Acoustics, Speech and Signal Processing (ICASSP)}},
  \bibinfo{pages}{5479--5483} (\bibinfo{organization}{IEEE},
  \bibinfo{year}{2018}).

\bibitem{holly2020profiling}
\bibinfo{author}{Holly, S.}, \bibinfo{author}{Wendt, A.} \&
  \bibinfo{author}{Lechner, M.}
\newblock \bibinfo{title}{Profiling energy consumption of deep neural networks
  on nvidia jetson nano}.
\newblock In \emph{\bibinfo{booktitle}{2020 11th International Green and
  Sustainable Computing Workshops (IGSC)}}, \bibinfo{pages}{1--6}
  (\bibinfo{organization}{IEEE}, \bibinfo{year}{2020}).

\bibitem{kaup2014powerpi}
\bibinfo{author}{Kaup, F.}, \bibinfo{author}{Gottschling, P.} \&
  \bibinfo{author}{Hausheer, D.}
\newblock \bibinfo{title}{Powerpi: Measuring and modeling the power consumption
  of the raspberry pi}.
\newblock In \emph{\bibinfo{booktitle}{39th Annual IEEE Conference on Local
  Computer Networks}}, \bibinfo{pages}{236--243} (\bibinfo{organization}{IEEE},
  \bibinfo{year}{2014}).

\end{thebibliography}



\section*{Acknowledgements}

SKT and MKK are partially supported by the EPSRC programme grant TRANSNET, EP/R035342/1. PJF and DAR acknowledge the support from the EU Horizon 2020 Marie Skodowska-Curie Action projects No. 813144 (REAL-NET) and 860360 (POST-DIGITAL), respectively. JEP and SKT  acknowledge the support of the Leverhulme Trust project RPG-2018-063. 

\section*{Author contributions statement}
DAR, PJF, and JEP conceived the study. DAR and PJF proposed the neural network model. DAR performed the numerical simulations, designed the experimental set-up and obtained the experimental results. PJF generated the data and performed the architecture optimization. DAR and PJF designed the figures and tables.
DAR, PJF, and JEP wrote the manuscript, with the assistance of MKK and SKT. All authors reviewed the manuscript. The work of DAR was supervised by MKK and SKT. The work of PJF was supervised by JEP, AN and SKT. 

\section*{Data availability}
Data underlying the results presented in this paper are not publicly available at this time, but can be obtained from the authors upon request.

\section*{Competing interests}
The authors declare no competing interests.

\section*{Additional information}
\textbf{Correspondence} and requests for materials should be addressed to D.A.R or S.K.T






\end{document}